\documentclass{aa}  
\usepackage{graphicx}
\usepackage{txfonts}
\usepackage{pdflscape}
\usepackage{hyperref}
\usepackage{colortbl}
\usepackage{float}
\usepackage{adjustbox}
\usepackage{lscape}
\usepackage{comment}
\usepackage{changepage}
\usepackage{geometry}
\usepackage{threeparttable}
\hypersetup{linkcolor=blue, citecolor=blue, colorlinks=true}

\usepackage[normalem]{ulem}

\begin{document} 

   \title{A new automated tool for the spectral classification of OB stars}

   \author{E. Kyritsis
          \inst{1,2} \thanks{ekyritsis@physics.uoc.gr},
          G. Maravelias \inst{3,2},
          A. Zezas \inst{1,2} ,
          P. Bonfini \inst{4,2},
          K. Kovlakas \inst{1,2},
          P. Reig \inst{2,1}
          }
    \authorrunning{Kyritsis et al.}
   \institute{Physics Department, \& Institute of Theoretical and Computational Physics,   University of Crete, GR 71003, Heraklion, Greece
              \and
              Institute of Astrophysics, Foundation for Research and Technology-Hellas, GR 71110 Heraklion, Greece
              \and
              Institute for Astronomy, Astrophysics, Space Applications and Remote Sensing, National Observatory of Athens, GR-15236, Penteli, Athens, Greece
              \and
              Computer Science Department, University of Crete, GR 71003, Heraklion, Greece
             }
  
  \date{Received XX XX, XXXX; accepted XX XX, XXXX}

\abstract{}{}{}{}{} 
  \abstract
  {As an increasing number of spectroscopic surveys become available, an automated approach to spectral classification becomes necessary. Due to the significance of the massive stars, it is of great importance to identify the phenomenological parameters of these stars (e.g., the spectral type), which can be used as proxies to their physical parameters (e.g., mass and temperature).}
  {In this work, we aim to use the random forest (RF) algorithm to develop a tool for the automated spectral classification of OB-type stars according to their sub-types.}
  {We used the regular RF algorithm, the probabilistic RF (PRF), which is an extension of RF that incorporates uncertainties, and we introduced the KDE - RF method which is a combination of the kernel-density estimation and the RF algorithm. We trained the algorithms on the equivalent width (EW) of characteristic absorption lines measured in high-quality spectra ( Signal-to-Noise (S/N) $(>\sim 50$) from large Galactic (LAMOST, GOSSS) and extragalactic surveys (2dF, VFTS) with available spectral types and luminosity classes. By following an adaptive binning approach, we grouped the labels of these data in 11 spectral classes within the O2-B9 range. We examined which of the characteristic spectral lines (features) are more important for the classification based on a number of feature selection methods, and we searched for the optimal hyperparameters of the classifiers to achieve the best performance.
}
  {From the feature-screening process, we find that the full set of 17 spectral lines is needed
to reach the maximum performance per spectral class. We find that the overall accuracy score is $\sim 70 \%,$ with similar results across all approaches. We apply our model in other observational data sets providing examples of the potential application of our classifier to real science cases. We find that it performs well for both single massive stars and for the companion massive stars in Be X-ray binaries, especially for data of similar quality to the training sample. In addition, we propose a reduced ten-features scheme that can be applied to large data sets with lower S/N $\sim 20-50$.}
  {The similarity in the performances of our models indicates the robustness and the reliability of the RF algorithm when it is used for the spectral classification of early-type stars. The score of $\sim 70 \%$ is high if we consider (a) the complexity of such multiclass classification problems (i.e., 11 classes), (b) the intrinsic scatter of the EW distributions within the examined spectral classes, and (c) the diversity of the training set since we use data obtained from different surveys with different observing strategies. In addition, the approach presented in this work is applicable to products from different surveys in terms of quality (e.g., different resolution) and different formats (e.g., absolute or normalized flux), while our classifier is agnostic to the luminosity class of a star, and, as much as possible, it is metallicity independent.}  


   \maketitle
%

\section{Introduction }\label{sec:1}
Massive stars, although rare, can be characterized as cosmic engines \citep{Maeder} since they play a leading role in the evolution of their host galaxies and the Universe in general. During their short but very interesting lives, as well as their extremely violent deaths, these stars are excellent laboratories for studying of a wide range of astrophysical phenomena relevant to many different aspects of the study of the Cosmos.

Massive stars spend the longest part of their life in the main sequence as OB-spectral-type stars, mainly emitting UV photons and forming \ion{H}{II} regions. Hence, they can be used as valuable tracers of star formation (e.g., \citealt{Steidel_1996,Preibisch}). Due to their strong stellar winds, they are useful probes of present-day element abundances in the Galaxy (e.g., \citealt{Gies,Przybilla}), but also in external galaxies beyond the Local Group (e.g., \citealt{Urbaneja,Castro}). Because of their high luminosities, massive stars are ideal stellar objects for the determination of extragalactic distances (e.g., \citealt{Kudritzki}). Through their violent deaths, massive stars are the progenitors of objects such as supernovae, black holes, and neutron stars (e.g., \citealt{Poelarends,Smith}). Furthermore, they can be the origin of phenomena such as gamma-ray bursts (GRBs) or the recently discovered gravitational wave sources produced by a merger of two compact objects (e.g., \citealt{Langer,Abbott}). Although, the evolution of massive stars is still unclear, besides their initial mass there are also other significant factors that can determine their fate such as mass loss \citep{Massey,Smartt,Vink}, rotation \citep{Muller,Lee}, metallicity \citep{Bouret}, and binarity \citep{Sana_2013,Sana_2017}.

Traditionally, spectroscopy is the basic tool for studying the nature of massive stars, because their spectrum reflects their physical parameters (e.g., temperature, gravity, and chemical abundances). Although the detailed analysis of a stellar spectrum using fitting techniques provides the most complete array of information, this approach is generally computationally intensive and it requires high-quality data. On the other hand, the knowledge of a phenomenological parameter, such as the spectral type, is very useful for statistical studies of large samples of massive stars. In particular, the spectral type and the luminosity class of a star are correlated with its physical parameters (e.g., temperature, mass, and radius). Based on this information, we can determine ages and formation scenarios for these systems.

Spectral classification is traditionally performed through the visual examination of a spectrum \citep{Morgan}. However, since the entire process is based on the presence or absence of diagnostic spectral lines, it is qualitative in nature and suffers from subjectivity, despite the power of the human eye as a pattern recognition classifier. Furthermore, it is an extremely time-consuming method, unable to handle a large volume of data. Nowadays, the continuously expanding volume of astronomical data sets, both in size and complexity, have commenced the era of big data in astronomy \citep{Pesenson}. In particular, big spectroscopic surveys for OB stars such as IACOB \citep{IACOB_2011_a}, NoMaDS \citep{NoMADS}, and GOSSS \citep{GOSS_sur}, have started producing a large volume of data, which is obviously extremely difficult to classify using traditional spectral classification. As a result, the best way to address these problems is the development of automated methods based on quantitative measurements of spectral features.

Significant work has been done in this direction, based either on the assessment of specific criteria (e.g., spectral features) or on pattern recognition (e.g., \citealt{vonHippel}). The former approach imitates what a human classifier actually does when visually examining a spectrum and estimating the presence of spectral lines. The latter is based on searching for the spectrum that best resembles the
observed one in a library of template spectra for different spectral types \citep{Duan,Scibelli}. Furthermore, there are works that combine photometric properties with measurements of spectral features (e.g., \citealt{Allende, Gkouvelis}). Even though these methods improve the automated spectral classification, they face some challenges. Criteria-evaluation techniques have difficulty accounting for all spectral features within a wide range of spectral types and their sub-types. On the other hand, template matching methods need observed data of similar quality with the templates in terms of wavelength range, resolution, and Signal-to-Noise (S/N). However, this is not typical for data obtained with different instruments and/or observational strategies. Furthermore, they are computationally intensive, limiting their use in large data sets.

As a result, the next choice for the automation of spectral classification is the use of machine-learning algorithms, which are divided into two groups: supervised and unsupervised. The former use known and labeled data as inputs, while the latter try to learn the various groups of the data from the data itself \citep[for a full overview, see][]{Baron}. Recently, thanks to improvements on software development and computational power, an increasing number of studies apply machine learning to a variety of problems in astronomy, taking advantage of its computational speed, as well as, its ability to handle large volumes of data (e.g., \citealt{Mahabal_2008,Laurino,Castro_2018,Pearson,Clarke,XRBs_classification}).

More specifically, a number of previous studies tried to develop classifiers for the spectral classification of stars. \cite{Navarro_2012} proposed a classifier for spectra with low S/N  based on artificial neural networks, while \cite{Mahdi} used probabilistic neural networks for the development of their classifier. In addition, \cite{Sharma} identified convolutional neural networks as the best-performing method among two supervised algorithms and one unsupervised one. Also, \cite{Kheirdastan} compared probabilistic neural networks and a K-means cluster, as well as a support vector machine to show that neural networks perform better than the others. Finally, \cite{Li} used a random forest algorithm and investigated the spectral feature evaluation problem. Although the aforementioned studies propose automated spectral classification methods, all of them present a more coarse classification from O-type to M-type spectral range without focusing on the OB stars. However, a finer spectral classification of these stars is very important since the difference on the spectral lines is significant among sub-spectral types because the temperature decreases drastically. 

In addition, some of these methods rely on Balmer lines.
This is a problem for Oe/Be stars, which account for up to $10-15\%$ of B-type stars \citep{McSwain_2005,Martayan_2007,Maravelias_2017}. These stars are particularly important since they are considered to occupy the upper-end of stellar rotation velocity distribution with velocities close to their breakup limit. The reason we cannot use Balmer lines for the classification of these stars is that the strength and profile of the lines varies due to the changing environments of these stars. In addition, Be stars are the donor stars in Be X-Ray Binaries (BeXBs), the main subcategory of high-mass X-ray binaries \citep[HMXBs;][]{Reig_review}. The assignment of a spectral type on the optical companions of these systems is very important. Based on the spectral-type--mass distributions of HMXBs, we can investigate differences between different populations (e.g., HMXBs in the Galaxy and the Magellanic Clouds) and understand how the star-formation history and/or metallicity can influence their formation rates \citep{Linden,Tzanavaris,Antoniou_SFH_LMC,Antoniou_SFH_SMC}. As a result, there is a need for the development of alternative classification schemes that do not rely on the Balmer series.
In order to overcome the limitation of using generic spectral classification tools in early-type stars, and particularly Oe/Be stars (which require special treatment), we developed a spectral classifier tailored to these very interesting stars.

This paper is organized as follows. In Section \ref{sec:2}, we describe the training sample and processing of the data. In Section \ref{sec:3}, we present an overview of the algorithms used in this work as well as their implementation and their optimization. In Section \ref{sec:4}, we present the performance of the algorithms, and in Section \ref{sec:5} we discuss these results and apply our model to a number of science cases. Finally, in Section \ref{sec:6} we summarize our conclusions and discuss possible future improvements. 

\section{Data collection and processing} \label{sec:2}
\subsection{Spectroscopic data} \label{subsec:samples} 
In this work, we used a collection of different spectroscopic surveys targeting the Galaxy and the Magellanic Clouds. We include the Large Magellanic Cloud (LMC; 1/2.5 of solar metallicity) and the Small Magellanic Cloud \citep[SMC; 1/5 of solar metallicity;][]{Russell_1992} in order to develop a model that can be applied to a wide range of metallicity environments. While each survey has a different wavelength coverage, their common range in the optical regime  (${\sim}3900-4900$ $\AA$) includes the most characteristic spectral lines of OB stars \citep{Walborn}. The reported spectral types in these surveys were obtained through different approaches (either visual inspection or via template fitting). For our purposes, we treated these spectral type labels as the ground truth. Furthermore, aiming as much as possible for a Luminosity Class (LC) independent model that can be applied for the analysis of survey data without any pre-processing, we took into account all the available LCs I-V per spectral type from each survey.

Our Galactic sample consists of spectra obtained from the Galactic O-Star Catalog \citep[GOSC\footnote{\url{http://ssg.iaa.es/}}][]{GOSS_catal}, which mainly includes stars from the Galactic O-Star Spectroscopic Survey \citep[GOSSS][]{GOSS_sur}. GOSSS is a massive spectroscopic survey of Galactic O-type stars based on high S/N ${\sim}$ 250, resolving power ($R{\sim}$2500), and blue-violet observations from both hemispheres. The observations for this survey were obtained from three different telescopes: the 1.5m at Observatorio de Sierra Nevada (OSN), the 2.5m duPont telescope at Las Campanas Observatory, and the 3.5m telescope at Calar Alto (CAHA). They have a spectral coverage of $\sim 3900-5100$ $\AA$.

Also, we obtained spectra from the catalog of OB stars from the Large Sky Area Multi-Object Fiber Spectroscopic Telescope \citep[LAMOST\footnote{\url{http://www.lamost.org}};][]{OB_catalog_lamost}. LAMOST is a reflecting four-meter Schmidt telescope, and its unique design has allowed it so far (DR5), to observe almost eight million Galactic stellar spectra with spectral coverage of ${\sim} 3700 - 9000$ $\AA$, S/N  between 20-300, and a resolving power of $R{\sim}$1800 \citep{LAMOST_1,LAMOST_2}.

Our extragalactic sample consists of spectra obtained from\\
the 2dF survey of the Small Magellanic Cloud, which is an extensive spectroscopic survey of O-, B-, and A-type stars in the ${\sim}3900-4800$ $\AA $ wavelength range, with a resolving power of $R{\sim}$1500 and S/N of ${\sim}$ 20-150 \citep{Evans}. The observations for this survey were obtained using the multiobject mode of the 2dF spectrograph that was mounted at the top end of the Anglo-Australian Telesope (AAT).

We also obtained spectra from the VLT-FLAMES Tarantula Survey (VFTS), which was a Large Programme at the European Southern Observatory (ESO). The VFTS obtained multiepoch optical spectroscopy of over 800 OB-type stars in the 30~Doradus region of the Large Magellanic Cloud (LMC). An overview of the observations is given by \cite{VFTS_paper_I}. The survey primarily used the Medusa--Giraffe mode of the Fibre Large Array Multi-Element Spectrograph (FLAMES) on the Very Large Telescope (VLT). Three of the standard Giraffe settings were used, giving coverage of 3960-5071\,\AA\ at $R$\,$\sim$\,7500 (LR02, LR03 settings) and 6442-6817\,\AA\ at $R$\,$=$\,16\,000 (HR15N setting). The spectra used here are those from the study by \cite{VFTS_paper_XVIII} that provided classifications and radial velocities of the B-type stars from the VFTS. In their study, the LR02 and LR03 spectra were co-added for the apparently single stars (and the single-lined binaries, see Sect.~2 of their paper for further details) and then smoothed and rebinned to an effective resolving power of $R$\,$\sim$\,4000 for classification; it is these classification B-type spectra that we used here.

For the construction of a uniform data set for the training and the evaluation of the algorithms, we handled each survey separately. For the GOSC data, we collected all the publicly available spectra in the wavelength range we are interested in ($\sim 3900-4900$ $\AA$), their spectral types, and their LCs. For a number of spectra, the LC information was not available, thus we decided to reject 17 out of 584 objects. For the 2dF data, we followed the same procedure as for GOSC resulting in the collection of 700 spectra. For the VFTS, we rejected seven spectra without known LCs out of 431 objects. In addition, we rejected 51 objects for which their spectral types had an uncertainty larger than 1 spectral type (i.e., B0.5-B2, B1-B3, B0.7-B1.5, B1-B2, and B2-B3) and one object with spectral type A0, as well as two objects without exact spectral types (classified as early B and mid-late B).

Finally, the entire LAMOST catalog of OB stars consists of 22901 spectra of 16.032 stars in the O9-F8 spectral-range and I-V luminosity class. Initially, for stars with multiple exposures we kept only the ones with the highest g-band S/N. Then, we applied a series of quality cuts (g-band S/N $>50$, removed objects with bad quality flags, missing fluxes). The remaining sample includes 5185 stars with spectral types OB, their sub-types and their LCs within the range I-V. On top of that, we applied an additional cut in order to reduce the contamination by A-type stars misclassified as B9 or B9.5, which is a caveat stated in the original catalog. In particular, we cross-correlated all B9-B9.5 stars with templates of B9 and A0 stars built by manually choosing two high-quality LAMOST spectra of each category. The cross-correlated spectra were first normalized by fitting a high-order polynomial to their continuum, and the template that resulted in the lowest $\mathbf{\chi^{2}}$ was considered as the most appropriate. This way, we identified 292 A0 stars within the B9-B9.5 stars. After removing them from the LAMOST sample, we continued with the remaining 4893 spectra.

\begin{figure}
        \includegraphics[width=\columnwidth]{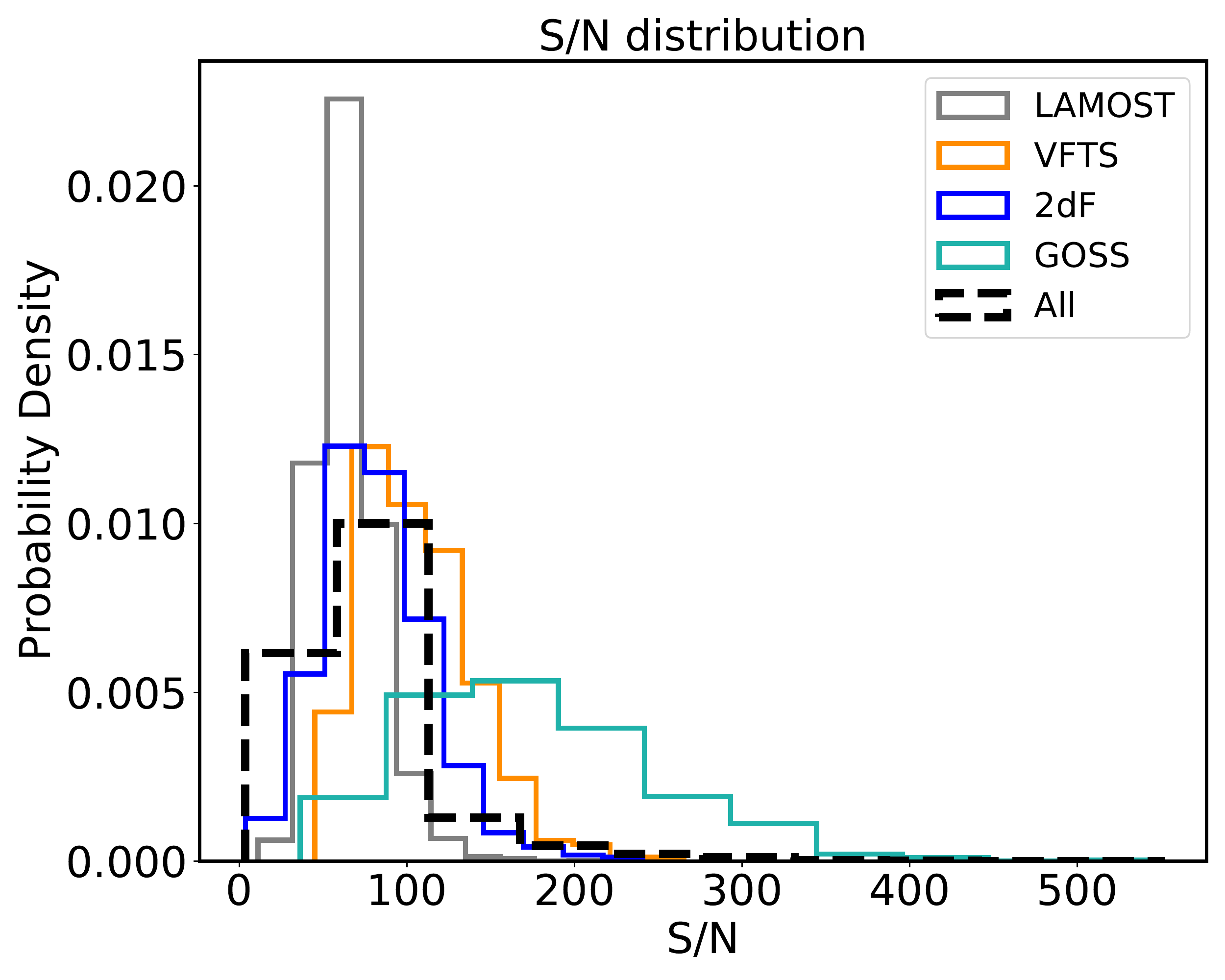}
    \caption{S/N distribution per survey as it is measured in a clear continuum region within the wavelength range 4220-4280 \AA. The statistical mode for “All” distribution is $\sim 42,$ and the $95^{th}$ percentile is $\sim 159$. Based on this result, we continued our analysis by only taking into account spectra with $S/N > 50$.}
    \label{Fig.:S_N_distribution}
\end{figure}

In order to construct a homogeneous sample in terms of S/N, we measured the S/N of each spectrum in exactly the same way. For this reason, by visually inspecting a large fraction of randomly selected spectra from all available surveys, we defined a continuum region within the 4220-4280 \AA \, wavelength range free of strong lines. Afterwards, we calculated the S/N for each spectrum by using the following formula: 
\begin{equation}
S/N = \frac{\Bar{F}_{4220-4280}}{\sigma_{F_{4220-4280}}}\,
,\end{equation}\label{S/N_formula}
where the $\Bar{F}_{4220-4280}$ is the mean flux of the spectrum within the selected spectral region and the $\sigma_{F_{4220-4280}}$ is the standard deviation of the flux in the same wavelength range. The S/N distribution per survey is presented in Fig. \ref{Fig.:S_N_distribution}.

The mode for the distribution of the S/N for the overall sample ('All') is $\sim 42$ and the $95^{th}$ percentile is $\sim 159$. Thus, we defined a S/N cut at 50 and we excluded spectra with S/N < 50 from our analysis. Despite it being higher than the mode, we chose this value as the minimum S/N of an input spectrum for two reasons: a) after visual inspection of a large number of the available spectra, we found that, in general, spectra with S/N $\sim$ 40 are not good enough to measure weak spectral lines; b) by choosing a S/N cut much higher than 50, we lose a large fraction of the available data for the training of our classifier, as is shown clearly in Fig. \ref{Fig.:S_N_distribution}. As a result, we opted for this specific S/N cut as a compromise between good quality spectra and adequate number of data.

After the previous S/N filtering, our final sample contains 5329 objects with the following demographics:
\begin{enumerate}
\item 549 galactic sources from the GOSC in a spectral type range between the O2-O9.7 types and LC I-V.
\item 3822 galactic sources from the LAMOST catalog in a spectral range between the O9-B9.5 types and LC I-V.
\item 590 SMC sources from the 2dF survey in a spectral range between the O5-B9 types and LC I-V.
\item 368 LMC sources from the VFTS in a spectral range between the B0-B9 types and LC I-V.
\end{enumerate}

Since our final sample encompasses spectra from both Galactic and extragalactic sources, we de-redshifted all of them to their rest-frame wavelength. The two extragalactic surveys, VFTS and 2dF, were corrected assuming the line-of-sight velocity of the LMC and the SMC from the SIMBAD database \citep{simbad}. The LAMOST spectra are already redshift-corrected, so we did not apply any further correction. Finally, for the GOSC spectra after visual inspection, we found that they are corrected for velocity shifts.

\subsection{Classification scheme} \label{subsec:features}
To build an appropriate spectral type classification scheme for our algorithm, we relied on the classification criteria for B-type stars in the SMC, as they were defined in the previous works of \cite{Evans} and \cite[][see Table 2]{Maravelias}. According to this scheme, we constructed a sample of 17 characteristic spectral lines by including both \ion{He}{I} lines, \ion{He}{II} lines (strong indicators for early type O-stars), and metal lines such as \ion{Mg}{II}, \ion{Si}{I}, \ion{Si}{II,} etc. (indicators of late type B-stars). In general, despite the fact that Balmer lines are widely used for the spectral classification of OB stars, we intentionally did not consider them in our analysis. Oe and Be stars and those which are found as companions in BeXBs create a circumestellar disk that can be detected observationally from the presence of strong Balmer emision lines. In particular, the variability of the disk's size and geometry (in timescales of months) results in a variable Balmer series \citep{Porter_Rivinious,Reig_review}. When the circumstellar disk is absent the Balmer lines are in absorption but when the circumstellar disk is fully developed the Balmer lines are in strong emission. As a result, we cannot take advantage of these lines since their presence does not depend only on the star's temperature but also on the disk's size. Finally, although Table 2 in \cite{Maravelias} includes the \ion{Ca}{II}K/3928 spectral line, we did not use it because the wavelength range of our surveys did not cover a large enough bluewards region for reliable continuum subtraction. However, this does not significantly affect the classification scheme because \ion{Ca}{II}K/3928 is not a strong discriminator between O and B spectral types in comparison with other spectral lines included in our classification scheme.

\subsection{Spectral type binning}
\begin{figure}
        \includegraphics[width=\columnwidth]{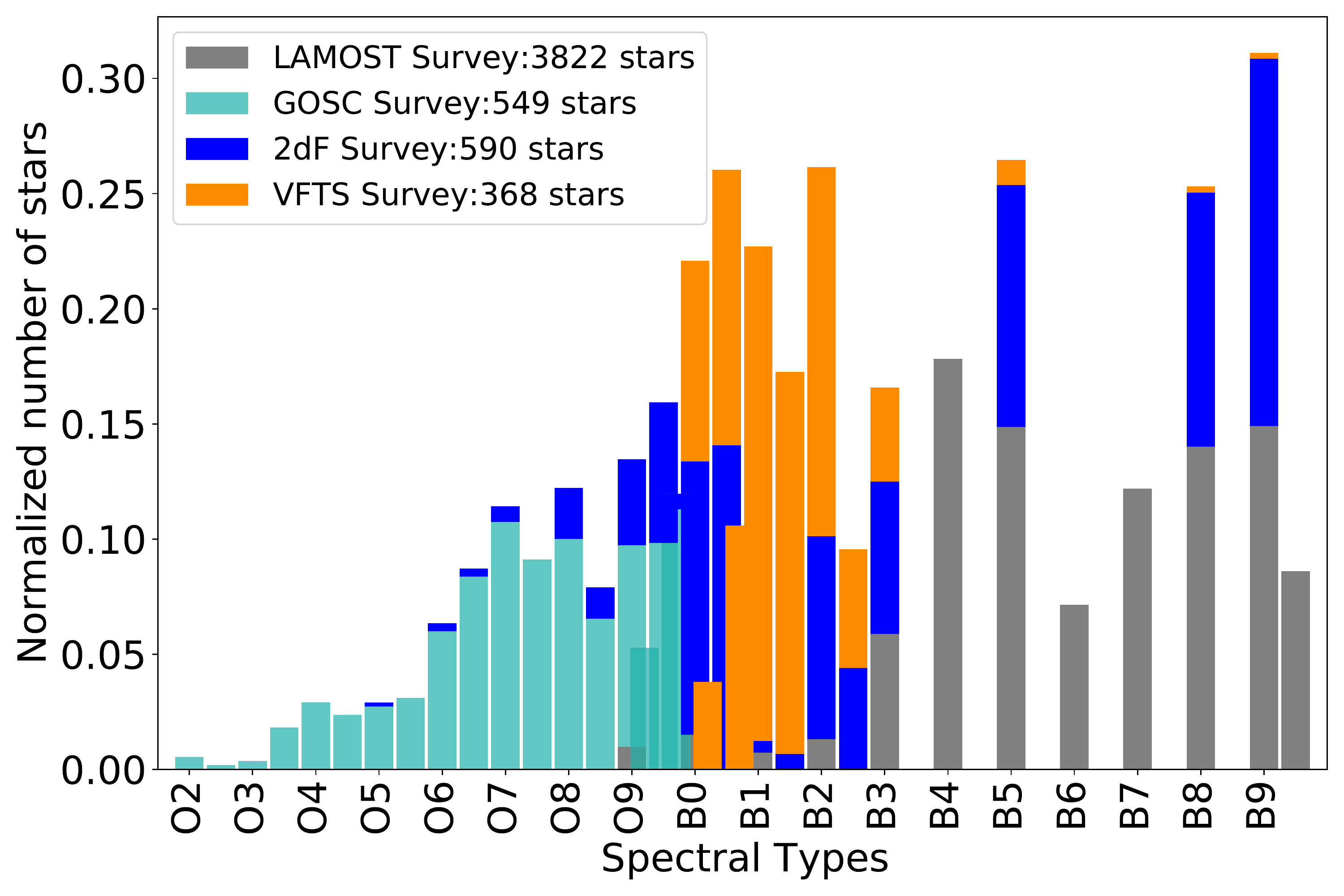}
        \includegraphics[width=\columnwidth]{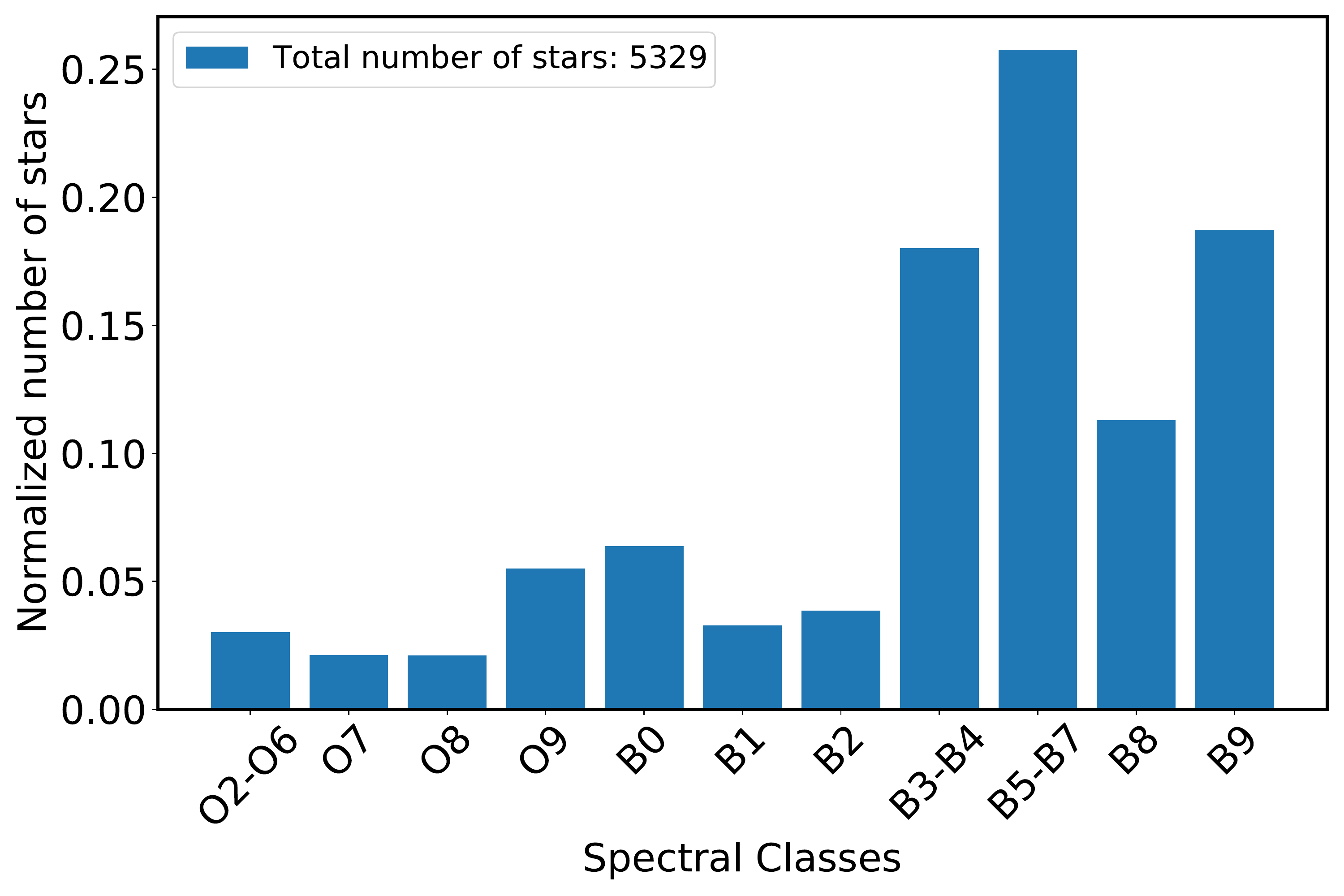}
    \caption{Top panel shows the initial spectral types distribution. The number of types is high enough, with some of them having a low number of objects. This distribution is not effective for training our algorithms, especially for the underrepresented classes. The bottom panel shows the final labels after the application of the adaptive binning. The number of spectral classes is decreased from 32 to 11.}
    \label{Fig.:initial_final_sample_distr}
\end{figure}

In the top panel of Fig. \ref{Fig.:initial_final_sample_distr}, we present the initial spectral type distribution of the stars in our sample. For a number of spectral types, our sample only included a few objects (e.g., O2, O3, O3.5 or O9.2,  and B0.2). These underrepresented types had fewer than 50 objects each, meaning that the machine learning algorithm cannot be trained efficiently. On the other hand, it is very hard to distinguish between adjacent spectral types of mid (B3-B5) or late (B5-B8) B-type stars 
\citep[see e.g., Table 2 of][]{Maravelias}.
Although in this case we have enough objects per class for the training of the algorithm, the available spectral lines cannot discriminate between neighboring spectral types \citep{Evans,Maravelias,Gray}. To overcome these limitations, we defined our classification classes following an adaptive binning scheme based on the resolution of our classification scheme, while ensuring that there are at least $\mathbf{} 100$ in each bin. The final scheme followed in our analysis is presented in Table \ref{tab:adaptive-binning}. This way, we reduced the number of spectral types examined from 32 to 11 without losing the physical meaning of the grouping (e.g., grouping all earliest types together) and we increased the number of objects even for the classes with a very small number of stars in our sample.
\begin{table}
\centering
\caption{The adaptively binned spectral classes considered in our analysis, and the original spectral types.}
\begin{tabular}{cc}
\hline\hline
Original spectral types & Grouped spectral classes\\\hline
O2-O6.5& O2-O6 \\
O7-O7.5& O7 \\
O8-O8.5  & O8 \\
O9-O9.7& O9 \\
B0-B0.7& B0\\
B1-B1.5& B1 \\
B2-B2.5& B2 \\
B3-B4& B3-B4 \\
B5-B7& B5-B7 \\
B8& B8 \\
B9-B9.5& B9 \\
\hline
\end{tabular}
\label{tab:adaptive-binning}
\end{table}
In the bottom panel of Fig. \ref{Fig.:initial_final_sample_distr}, we present the final spectral class distribution of our sample as it was formed after the adaptive binning.

\subsection{EW and uncertainty measurements}\label{subsec:EWs_EWs_errors}
To quantify the intensity of the selected features, we measured the EW of each spectral line from the \emph{\emph{one-dimensional}} spectrum of each object in our data set. The EW is defined as the width of the continuum region of a spectrum that contains the same flux as the spectral line examined, and it is given by
\begin{equation}
EW = - \int_{\lambda_{1}}^{\lambda_{2}} \frac{F_{cont}(\lambda)-F_{line}(\lambda)}{F_{cont}(\lambda)}d\lambda = (\lambda_{1}-\lambda_{2})-\int_{\lambda_{1}}^{\lambda_{2}}\frac{F_{line}(\lambda)}{F_{cont}(\lambda)}d\lambda \ ,
\end{equation}\label{EW_basic_formula}
where $\lambda_{1}$and $\lambda_{2}$ are the initial and final wavelength over which the line flux is calculated, and $F_{cont}$ and $F_{line}$ are the continuum and spectral-line flux density, respectively.

To measure the EW, we used the \texttt{Sherpa fitting package v 4.13.0 } \citep{Sherpa} to model the characteristic spectral lines and their local continuum regions by fitting a Gaussian line and a polynomial, respectively. \texttt{Sherpa} is a very useful tool that allows us to measure the flux and the amplitude either of absorption or emission spectral lines using complex models, and to determine their parameters and their corresponding uncertainties while accounting for uncertainties on the data.

First, we visually inspected~100 randomly selected spectra of different spectral types and LC from each survey to define the regions used for the spectral fit, taking care to excise strong spectral features from the continuum bands. We decided to use slightly different continuum bands for the VFTS, LAMOST, and GOSC surveys to avoid spectral features at the edge of the continuum bands, which can be particularly strong in the higher resolution spectra. For the 2dF survey, we used the same bands as those used for the LAMOST survey. In Table \ref{sec:features_table}, we present the full scheme of the selected spectral lines with their central wavelengths and their corresponding spectral ranges, as well as the continuum regions that were used in the fitting process.

For the VFTS, 2dF, and GOSC surveys in which the spectra are normalized and they did not provide uncertainties, we adopted the standard deviation of the data in a relatively clear continuum region in the 4220-4280 $\mathbf{\AA}$ wavelength range as a measurement error. The continuum of these spectra was fit with a constant. On the other hand, for the unnormalized LAMOST spectra, which also provided measurement errors, we modeled the continuum with a$^{}$ first-order polynomial, and we used the values provided by the survey as flux errors. We fit each spectral line separately except for the \ion{Si}{IV}/4088\,$\mathbf{\AA}$ , \ion{Si}{IV}/4116\,$\mathbf{\AA}$ , \ion{He}{I}/4121\,$\mathbf{\AA}$, and \ion{Si}{IV}/4130\,$\mathbf{\AA}$ spectral lines. Since these lines are very close to \ion{H}{$\mathbf{\delta}$} line, resulting in strong blending particularly for late-type B-stars, they were fitted in a single model also including the \ion{H}{$\mathbf{\delta}$} line. In order to account for the strong wings of the \ion{H}{$\mathbf{\delta}$} line, we used two Gaussian components in the later type stars, one accounting for the narrow core of the line and one for its broad wing. The position of each line was at its rest-frame wavelength $\pm2$ \,\AA \, in order to avoid confusion with other neighboring lines (e.g., the \ion{Si}{III}/4553\,$\mathbf{\AA}$ and the \ion{Si}{III}/4550\,$\mathbf{\AA}$ lines). The final model for this spectral region consisted of four Gaussian lines, one for each spectral feature of interest, two Gaussians for the \ion{H}{$\mathbf{\delta}$} line, and a$^{}$ zeroth- or$^{}$ first-order polynomial for the continuum. To estimate of the model parameter uncertainties, we used the \texttt{covar} tool, which computes confidence intervals for the specified model parameters in the data set using the covariance matrix of the fit. The entire fitting process was based on an automated pipeline developed for this purpose.

\begin{table*} 
    \centering
        \caption{ Spectral lines with their wavelength ranges, their central wavelengths and continuum regions that have been used for the calculation of their EWs, during the spectral line fitting. We used slightly different continuum bands for VFTS, LAMOST and GOSC surveys because the different spectral resolutions of each of them can cause the presence of undesirable spectral features within the predefined continuum regions, adding noise in the data. Here, we present the continuum regions that selected for LAMOST data since their population dominates among the other surveys. This is the final scheme of features that is used through this work, except for the \ion{H}{$\delta$}/4100 $\mathbf{\AA}$ spectral line, which is used only for the fit.}
            \begin{tabular}{lccccccc}
            \hline\hline
            Line ID & \multicolumn{3}{c}{Spectral line} &
            \multicolumn{2}{c}{Continuum blue}  & \multicolumn{2}{c}{Continuum red} \\
            &$\qquad\lambda_{\textrm{central}}$ &  $\lambda_{\textrm{start}}$ & 
            $\lambda_{\textrm{end}}$  &  $\lambda_{\textrm{start}}$ & 
            $\lambda_{\textrm{end}}$  &  $\lambda_{\textrm{start}}$ & 
            $\lambda_{\textrm{end}}$ \\
            $\qquad$&({\AA}) & ({\AA}) & ({\AA}) & ({\AA}) & ({\AA}) & ({\AA}) &
            ({\AA}) \\
            \hline
            \ion{He}{I}   &4009 & 4004 & 4016 & 3990 & 4002 & 4035 & 4060 \\
            \ion{He}{I}+\ion{He}{II}   &4026 & 4017 & 4034 & 3990 & 4002 & 4035 & 4060 \\
            \ion{Si}{IV}  &4088 & 4084 & 4091 & 4055 & 4075 & 4150 & 4170 \\
            \ion{H}{$\delta$}  &4100 & 4095 & 4110 & 4055 & 4075 & 4150 & 4170 \\
            \ion{Si}{IV}  &4116 & 4113 & 4118 & 4055 & 4075 & 4150 & 4170 \\
            \ion{He}{I}   &4121 & 4118 & 4125 & 4035 & 4060 & 4150 & 4170 \\
            \ion{Si}{II}  &4130 & 4125 & 4135 & 4035 & 4060 & 4150 & 4190 \\
            \ion{He}{I}   &4144 & 4140 & 4150 & 4035 & 4060 & 4150 & 4170 \\
            \ion{He}{II}  &4200 & 4190 & 4207 & 4150 & 4190 & 4245 & 4260 \\
            \ion{Fe}{II}  &4233 & 4229 & 4237 & 4205 & 4225 & 4238 & 4260 \\
            \ion{He}{I}   &4387 & 4382 & 4392 & 4365 & 4380 & 4398 & 4415 \\
            \ion{O}{II}   &4416 & 4412 & 4421 & 4398 & 4411 & 4440 & 4460 \\
            \ion{He}{I}   &4471 & 4462 & 4477 & 4440 & 4460 & 4495 & 4535 \\    
            \ion{Mg}{II}  &4481 & 4477 & 4488 & 4440 & 4460 & 4490 & 4505 \\
            \ion{He}{II}  &4541 & 4537 & 4547 & 4510 & 4535 & 4590 & 4620 \\
            \ion{Si}{III} &4553 & 4548 & 4558 & 4485 & 4507 & 4600 & 4620 \\
            \ion{O}{II}+\ion{C}{III} &4645 & 4635 & 4655 & 4600 & 4625 & 4660 & 4685 \\
            \ion{He}{II}  &4686 & 4679 & 4692 & 4660 & 4670 & 4730 & 4745 \\
            \hline
            \end{tabular}
            \label{sec:features_table}
\end{table*}

After fitting each spectral feature of interest, we calculated the EW using the \texttt{eqwidth} routine provided by \texttt{Sherpa}. This tool calculated the distribution of EW for each spectral line by evaluating the line and the continuum flux based on drawings of the model parameters from the fit covariance matrix. We performed the calculation for 1000 draws and we adopted the median of the EW distribution and its $68 \%$ percentile as its uncertainty.

\begin{figure*}
        \includegraphics[width=2\columnwidth]{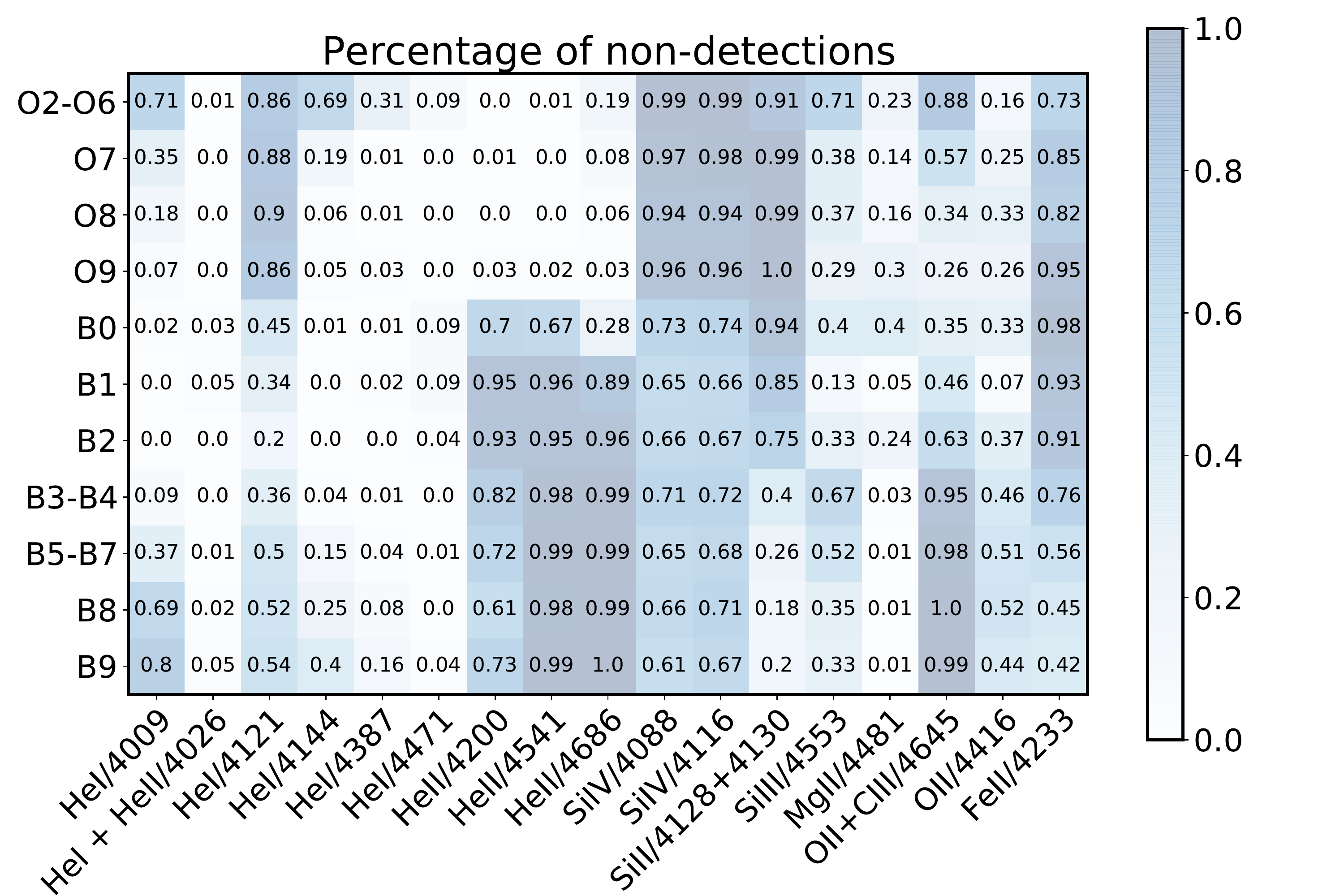}
        \caption{ Fraction of the non-detections per spectral type and per spectral line.}
    \label{Fig:non-detections}
\end{figure*}

Since in many cases some of the spectral lines were not detected, after the completion of the spectral fit we applied a post-processing step to identify non-detections. With regard to non-detections, we characterized the fits that had failed because of unconstrained parameters and all successfully measured spectral lines with an EW S/N \textless\ 3 $\left(\frac{EW}{\delta EW} < 3\right)$ or EW > 0 (emission lines). Instead of removing these cases from the training of our classifier, we included this information in our analysis by replacing their EW with the same constant (magic number) which had a different value for each line. In Fig. \ref{Fig:non-detections}, we present the fraction of non-detections per spectral type and per spectral line. This way, we do not reduce the size of our sample by removing a large number of objects with a few problematic lines, while at the same time we do include the diagnostically important information of the non-detection of some lines in our analysis. In addition, by selecting each magic number to be relatively far from the EW distribution of the successfully measured spectral lines, we ensure that the actual detections and non-detections do not overlap.\\

\section{Building the machine-learning models}\label{sec:3}
\subsection{Algorithms}\label{subsec:algorithms}
In this work, we used the widely used supervised learning RF algorithm \citep[for a thorough review, see][]{RF_review}. RF is a well-known ensemble method used for both classification and regression tasks (e.g., \citealt{Carliles,Vasconcellos}). Ensemble methods combine either different supervised learning algorithms or the information of a single algorithm that was trained on different subsets of the training set. The building blocks of the RF algorithm are the decision trees. A decision tree is a nonparametric model built during the training stage that describes the relation between the features and the labels using a set of continuous nodes in a tree-like structure. The initial node of a tree is called the root node, and the final nodes, in which the predicted label is calculated, are called terminal nodes. The input training data follow a top-to-bottom flow through the tree, which is split into children nodes according to the condition at each node. This condition is defined as the feature and its corresponding value that maximizes the class separation between the children nodes. This 'best' splitting threshold can be computed by using different metrics such as the Gini impurity, entropy, or information gain, with the most common choice being the first one. In a two-class classification problem (can be generalized to multiclass problems), the Gini impurity is defined as the probability of a randomly selected object to be misclassified if it is assigned with a label that is randomly drawn from the distribution of the labels in the group \citep{PRF}.

When we try to classify an unseen object with a decision tree, the object will be propagated through the tree, and, based on the values of its features and the conditions in the nodes, it will reach a terminal node with a specific label assigned to the object. However, the prediction power of a unique decision tree is limited since it is vulnerable to overfitting and cannot be generalized well to unseen data sets \citep{Breiman}. Solving for that, the RF algorithm constructs a number of decision trees, and during the training process of these trees it uses randomly selected subsets of the full data set based
on a bootstrap method. Furthermore, during this process, random subsets of the features are used in each node of each decision tree to find the appropriate conditions in the nodes to build each tree (i.e., determine its nodes). The final prediction of the RF algorithm is in the form of a majority vote. Each individual tree in the forest suggests a class for the examined object, and the final prediction is the class that has been proposed from the majority of the trees.

The RF algorithm has numerous advantages in comparison with other supervised algorithms. Firstly, due to the randomness on which it is built, the correlation between different trees is reduced. Hence, the structure and the conditions in the nodes of each tree are very different from tree to tree, resulting in a model that can be generalized well to new unseen data and consequently lead to improved performance. In addition, it can handle either categorical or numerical features with no need to normalize or scale the data, and it is effective in multiclass problems such as ours. On top of that, it can properly handle the non-detections that are fed to it through the magic number since the RF algorithm only needs a value for each feature in order to make a decision at each node. However, the most important benefit of the RF algorithm for our goal is that its tree-based logic resembles the traditional spectral classification technique.

Nevertheless, the RF algorithm is unable to take into account uncertainties of the data during the training process. To overcome this limitation, an alternative PRF approach was proposed by \cite{PRF}. The input of the PRF algorithm at each node is a quadruplet that contains the label and its corresponding uncertainty, alongside with the value of a feature and its uncertainty. The PRF treats these values as random variables, described by a normal probability distribution. In particular, the features become probability density functions (PDFs) with the mean value being the value of the feature and the variance being the square of the feature uncertainty, whereas the labels become probability mass functions (PMFs), where each label is assigned to an object with some probability. Since the spectral types provided with our data set do not include uncertainties for all surveys, we used the PRF accounting only for the uncertainties in the features. In the ideal case of negligible uncertainties, the PRF converges to the original RF.

We used the implementation of the RF classifier \texttt{sklearn.ensemble.RandomForestClassifier()} provided by the \texttt{scikit-learn}\footnote{\url{https://scikit-learn.org/stable/}} version 0.23.2 \citep{scikit-learn} package for Python 3. For the PRF we used the publicly available package at GitHub repository \footnote{\url{https://github.com/ireis/PRF}}. After processing the data described in Section \ref{sec:2}, we randomly shuffled our complete data set and split it into training and test data sets. We considered 70\% of the full set (3730/5329 spectra) for
the training data and 30\% (1599/5329 spectra) for the testing data. This is a common training/test ratio in machine learning that avoids overfitting \citep{Ksoll}.

\textbf{\subsection{Hyperparameter optimization}
\label{subsec:hyper-opt}}
A crucial step in building a machine learning model is the selection of appropriate values for the hyperparameters. Hyperparameters are those that control the learning process, and their values are set at the initialization of the training process. An inappropriate choice of these values can lead to underfitting or overfitting of the model. The RF algorithm includes a large number of hyperparameters. We chose to optimize our model for the most important ones, which are:
\begin{itemize}
\item \texttt{n\_estimators}: the number of trees in the forest.
\item \texttt{{max\_depth}}: the maximum number of levels in each decision tree.
\item \texttt{min\_samples\_split}: the minimum number of samples required to split an internal node.
\item \texttt{min\_samples\_leaf}: the minimum number of samples required to be at a leaf node.
\item \texttt{max\_leaf\_nodes}: this hyperparameter affects the way that trees grow in order to have the best result. The best nodes are defined as showing a relative reduction in impurity.
\item \texttt{max\_samples}: the number of samples to consider when the algorithm draws from the training data set to train each base estimator. 
\end{itemize}

\begin{figure}
        \includegraphics[width=\columnwidth]{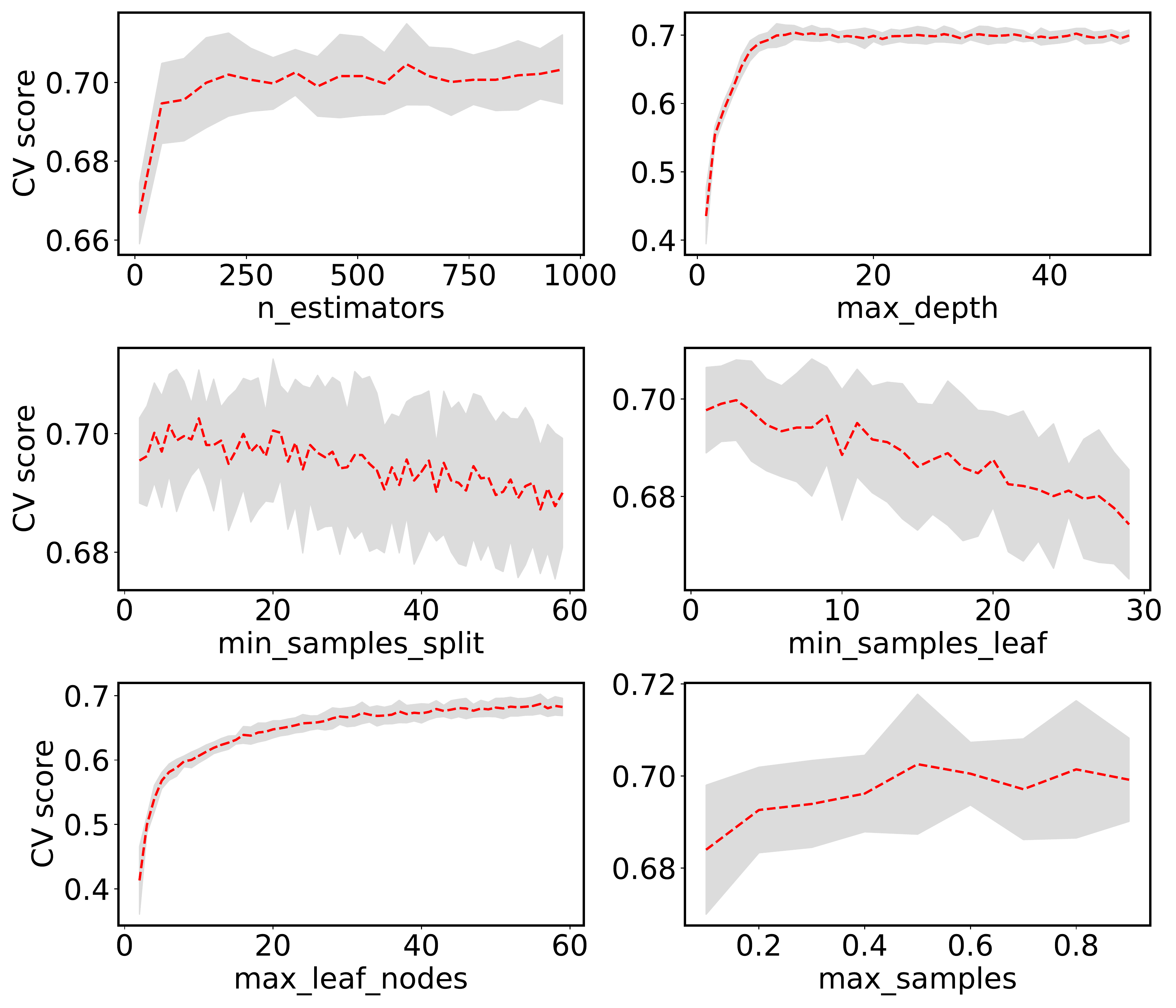}
        \caption{Validation curves of the most important RF hyperparameters, including the standard deviation of each CV score (gray shaded region). Based on these curves we selected the optimal values in the way described in Section \ref{subsec:hyper-opt}. The final values of the hyperparameters are tabulated in Table \ref{tab:opt_hyperpar}.}
    \label{Fig.:valid_curves}
\end{figure}

To optimize these hyperparameters, we trained and evaluated the RF model over a range of values (with others fixed to their default). After each iteration, we obtained the corresponding cross-validation accuracy score.The \textit{\emph{k-fold}} cross-validation (CV) accuracy score or rotation estimator \citep{Kohavi} is a technique used in order to take into account the entire data set and not only a part of it (when splitting into a training and test subsets). The initial data set is split into \textit{k} smaller data sets ("folds"). For each of the \textit{k}-folds, a model is trained using \textit{k-1} of the folds as a training set based on a stratified method (i.e., preserving the fractional representation of all classes in the original sample). Then, the resulting model is evaluated on the remaining data and reports an accuracy value. The final cross-validation score is the average over the folds and its uncertainty is the standard deviation of these values. In our work we, chose to use a \textit{k=5} for the number of folds. In Fig. \ref{Fig.:valid_curves}, we plot the validation curve for each hyperparameter; that is, the CV accuracy score for different values of the accuracy for the hyperparameter, as well as its standard deviation.

The \texttt{n\_estimators} increases quickly and beyond a threshold it remains almost constant. The same behavior is seen for the \texttt{max\_depth} and the \texttt{max\_samples}. In contrast, the \texttt{min\_samples\_leaf} and the \texttt{min\_samples\_split} show higher scores for lower values (close to their default), while gradually their scores decrease for larger values. Finally, the \texttt{max\_leaf\_nodes} seems to be consistent with its default value (\textit{None}), which means that the higher the number of leaf nodes is, the higher the accuracy of the model. The RF algorithm has two more important hyperparameters: \texttt{class\_weight} and \texttt{max\_features}. The former is useful for imbalanced data sets, while the latter controls the maximum number of features to consider when the algorithm is looking for the best split. We set the \texttt{class\_weight} hyperparameter to "\emph{\textit{\textit{balanced\_subsample}}}", which means that during the training of the algorithm, the weight per class will be inversely proportional to the number of objects belonging to this class. Moreover, \texttt{max\_features} will be equal to $\sqrt{N_{features}}$, where $N_{features}$ is the total number of features, and this is commonly used as default value for the RF algorithm.

In the previous test, when a range of values of a hyperparameter is tested all others remain constant. To check if there are any other combinations, we applied a grid search technique, which finds the best set of hyperparameter values trying all possible combinations over a predefined range of values. The model for each combination of values is evaluated based on k-fold CV test. It is a powerful and accurate method for tuning the hyperparameters but it is expensive in terms of computational time. Consequently, to reduce the computational time, we applied the grid search method using the range of values selected by the previous validation test. For \texttt{n\_estimators} and \texttt{max\_depth,} we selected the region where the algorithm's performance becomes constant. For \texttt{min\_samples\_leaf} and \texttt{min\_samples\_split,} we took into account a smaller region close to their default values since the validation curve indicated that the score remains constant at its maximum value over a very broad parameter range. Finally, for the \texttt{max\_samples} we considered all the values within 0.1-0.9. Furthermore, we did not consider the hyperparameter \texttt{max\_leaf\_nodes} for further analysis with the grid search method. This is because, from the corresponding validation curve, it is obvious that the default value is the optimal one. The results of the grid search are presented in Table \ref{tab:opt_hyperpar}.

Since the PRF algorithm is based on the RF algorithm, its most important hyperparameters are equivalent to the regular RF algorithm. These are \texttt{n\_estimators},  \texttt{max\_depth,} and  \texttt{max\_features}. In addition, the PRF includes one more hyperparameter called \texttt{keep\_proba,} which is the threshold of the propagation probability in the nodes, below which the nodes are pruned from the tree. This pruning reduces the computational time of the algorithm, without affecting its performance. By using the validation curves for these hyperparameters again, we saw that they exhibit similar behavior to that in the regular RF. As a result, we did not apply an additional grid search for them, and we considered exactly the same values as those derived from the RF hyperparameters' optimization. For \texttt{keep\_proba,} we set the value of 5\%, which is an optimal value according to \cite{PRF}, but it is also consistent with the result of its validation curve. In Table \ref{tab:opt_hyperpar}, we present the optimal values (last column) as they were derived from the hyperparameters' optimization for both RF and PRF models. In addition, the range of hyperparameter values and the number of steps that were considered during tuning of both models are tabulated. The optimal values in this table are the ones used to construct our final working models.

\begin{table*} 
    \centering
    \caption {Optimal hyperparameter values as they were derived after the optimization procedure for both RF and PRF models, as well as the value ranges and steps in which the optimal value was searched for. The optimal values define the final models used throughout this work.}
    \begin{tabular}{cccccc} 
    \hline
    \multicolumn{6}{c}{RF model} \\
    \hline
    Hyperparameter&Validation curve range&Step&Grid search range& Step &Optimal value\\
    \hline 
    n\_estimators               &10-950 &50  &100-950 &50&450\\
    max\_depth                  &1-49    &1   &10-49    &1 &15 \\
    min\_samples\_split         &2-59    &1   &2-9     &1 &5\\
    min\_samples\_leaf          &1-29    &1   &1-9     &1 &4\\
    max\_leaf\_nodes            &2-59    &1   &-        &- & \lq{default}\rq\\
    max\_samples                &0.1-0.9    &0.1   &0.1-0.9     &0.1 &0.6\\
    max\_features               &-    &-   &-     &- &\lq{auto}\rq\\
    class\_weight               &-       &-   &-        &- &\lq{balanced\_subsample}\rq\\
    \hline\hline
    \\
    \multicolumn{6}{c}{PRF model} \\
    \hline
    Hyperparameter&Validation curve range&Step& & &Optimal value\\
    \hline 
    n\_estimators               &10-950 &50 & & &450\\
    max\_depth                  &1-49    &1   & & &15\\
    keep\_proba                 &0-1    &0.01 & & &0.05\\
    max\_features               &-    &-   & & &\lq{auto}\rq\\
    \hline
    \end{tabular}
    \label{tab:opt_hyperpar}
\end{table*}
\textbf{\subsection{Feature selection}
\label{subsec:features-opt}}
Besides the optimization of the most important hyperparameters, another way to improve a model's performance is to investigate if there is any specific combination of features that can result in a better score. For this reason we used a sequential feature selection (SFS) algorithm in order to assess if we can reduce the error of the model when it is applied in other samples by removing irrelevant features that can introduce noise.

SFS algorithms are a family of search algorithms that are used to identify the minimal set of features that provide optimal information of a model. They are based on an iterative process, where the model's performance is evaluated for different feature subsets, the size of which is predefined by the user. For our analysis, we used the sequential forward floating selection (SFFS) \citep[see][]{Pudil} provided by the \texttt{Mlxtend} \citep[Machine learning extensions;][]{Raschka} library for Python. 

\begin{figure}
        \includegraphics[width=\columnwidth]{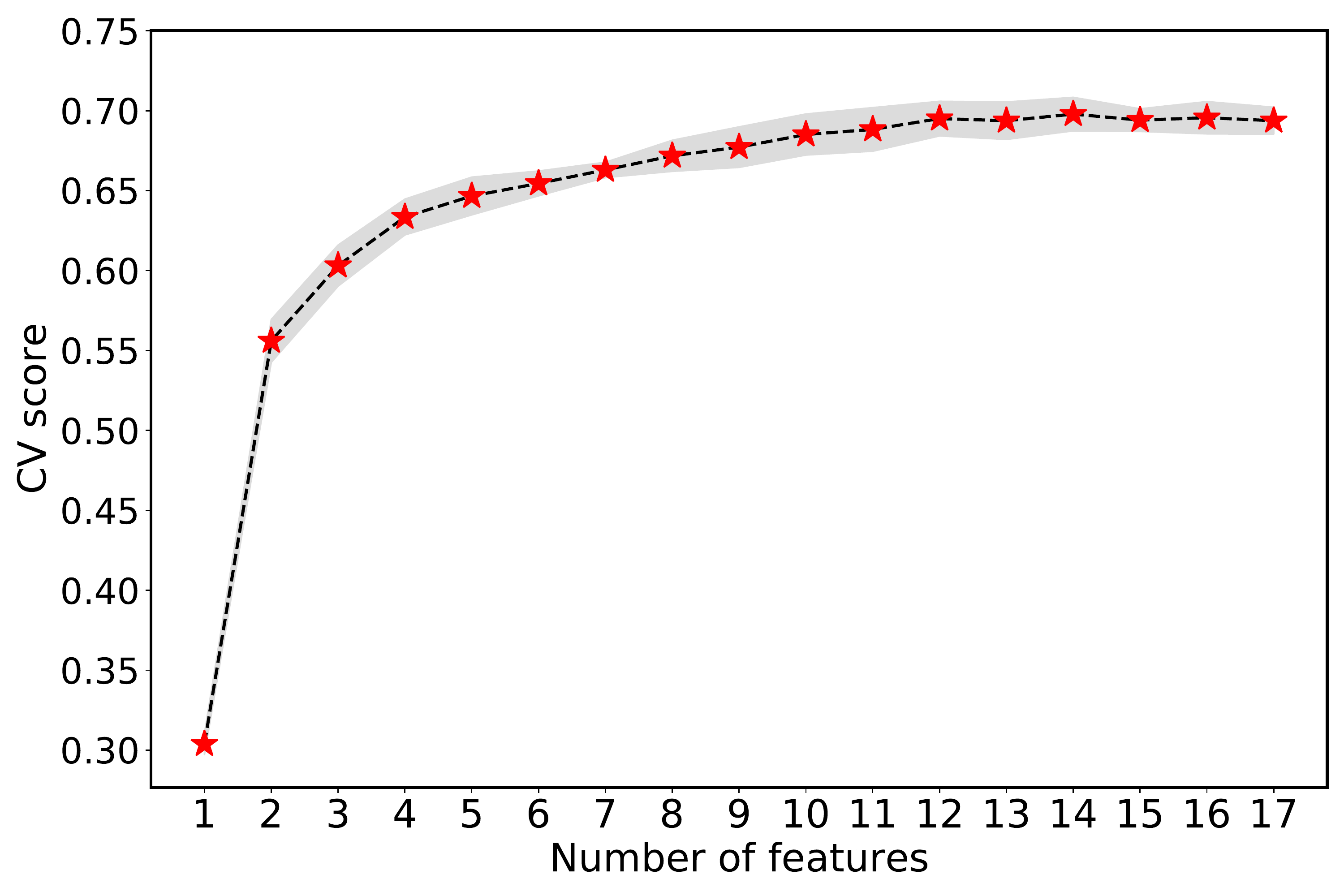}
        \caption{Number of features versus the CV accuracy score of RF algorithm. As it is shown the SFFS algorithm suggests that after ten features the model's performance does not change significantly. However, we considered the entire set of spectral lines (17) since even the weaker lines can improve the performance of particular spectral classes.}
   \label{Fig.:best_features}
\end{figure}

Using the above algorithm, we tested the full set of features in Table \ref{sec:features_table}, except for the \ion{H}{$\delta$} spectral line. Due to the similarity of RF and PRF, we ran the SFFS only for the best RF configuration, as determined from the hyperparameters' optimization. Since a priori we do not know the optimal size of the best feature subset, we added an extra step. We repeatedly executed the SFFS algorithm for all possible sizes, namely within the $1-17$range. Then, we plotted the size of these subsets; that is, the number of the features versus the CV score (based on the accuracy metric) that was suggested from the RF model evaluation, and we present it in Fig. \ref{Fig.:best_features}.
 
As shown in Fig. \ref{Fig.:best_features}, the SFFS algorithm suggests that after ten features the k-fold accuracy of the model remains constant, which means that any other addition of features does not change the model's performance. Furthermore, for a small number of features, the accuracy is low, which is expected, since the model has not enough information to classify correctly the objects in each class. We see that after ten features the CV score already reaches a value of 0.68, indicating that the classifier performs well even with ten features. Although this ten-feature scheme is comprised of the strongest spectral lines (e.g., \ion{He}{I},\ion{He}{II}) and the addition of extra features improves the CV score only by a factor $\sim 2 \%$, we do not consider it as the main classification scheme. Instead we adopted the entire classification scheme (17 spectral lines). This choice is driven, from an astrophysical perspective, by the fact that for distinguishing individual spectral classes the examination of specific, weaker lines is needed. As a result, the adoption of a classification scheme with only the strongest lines would cause our classifier to underperform in  spectral classes that rely on those weaker lines (e.g., the B8.0 class relies on the comparison of the \ion{He}{I} lines with the weak \ion{Si}{II}/4128+4130 \citep[see Table 2 in][]{Maravelias}. However, in Appendix \ref{10-feature scheme} we further discuss this ten-feature scheme as an alternative approach that can be applied to large data sets with lower S/Ns ($\sim 20 - 50$), where the measurement of weaker lines is not reliable.

\begin{figure*}
\centering
        \includegraphics[width=\textwidth]{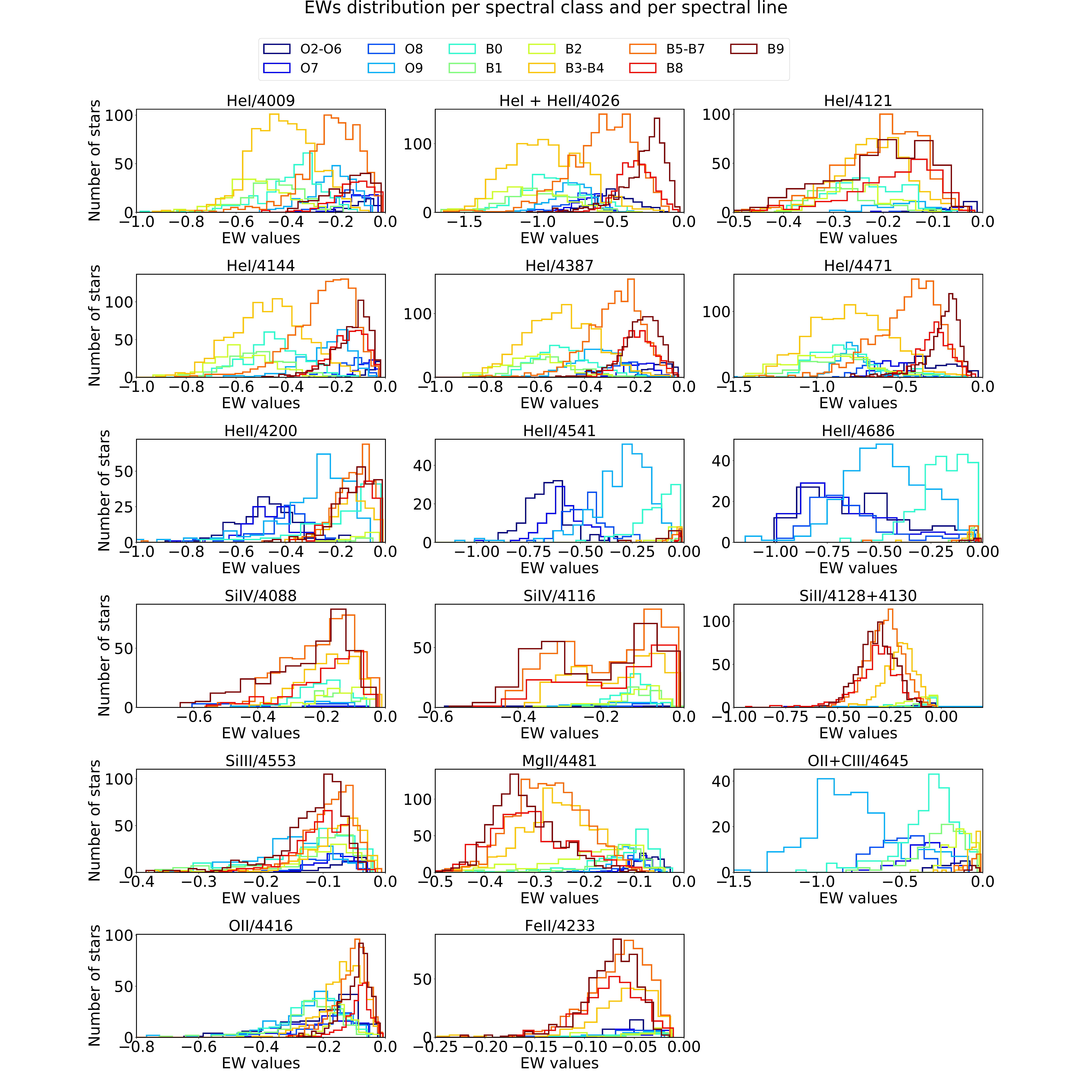}
        \caption{Distribution of EW values per spectral class and per spectral line for the full set of features as it is presented in Table \ref{sec:features_table}. The best-separated distributions of spectral lines coincide with the result of SFFS analysis (see Section \ref{subsec:features-opt}). Based on these distributions, as well as the result of the SFFS analysis, we constructed the final scheme of features that it is used throughout this work (Table \ref{sec:features_table}).}
    \label{Fig.:ew_distribs}
\end{figure*}

 To demonstrate that we can trust the adopted classification scheme with 17 features and understand the behavior of the SFFS algorithm, in Fig.\ref{Fig.:ew_distribs} we plot the EW distribution per spectral line and per spectral class only for the well-measured spectral lines (detections). From our experience with optical spectra, we expect that the SFFS algorithm will consider the spectral lines with well-separated EW distributions between different spectral classes as the most
useful features. In contrast, spectral lines with strong overlap between their EW distributions are expected as being less useful during the training of the algorithm. Furthermore, the \ion{He}{II} lines, which are present only in hot O stars, are expected to be distributed around values lower than 0, while for later type stars their EW values will be closer to zero; or, when the line is non-detected they will be replaced by the corresponding magic number. On the other hand, the distributions of lines such as the \ion{He}{I}/4471 $\mathbf{\AA}$ and the \ion{Mg}{II}/4481 $\mathbf{\AA}$ will show an anticorrelation with the former decreasing from the early-B to late-B, and the latter increasing from early-B to late-B. Indeed, a comparison between Fig. \ref{Fig.:ew_distribs} and the SFFS results (Fig. \ref{Fig.:best_features}) confirms that the majority of the spectral lines within our classification scheme have well-separated values per spectral class (e.g., \ion{He}{II}/4686 $\AA$, \ion{He}{I}/4026 $\AA$, \ion{O}{II}+\ion{C}{III}/4645 $\AA,$ or \ion{Mg}{II}/4481 $\AA$, etc.), and at the same time these lines are those that the SFFS algorithm considers as important (i.e., before it reaches its plateau). In contrast, the weaker spectral lines such as \ion{Fe}{II}/4233 $\mathbf{\AA}$ and \ion{Si}{IV}/4088 and \ion{Si}{IV}/4116 $\mathbf{\AA}$ provide no significant contribution to the algorithm's performance (incorporated in the model after the 10 features) due to the low separability of their EW values.
 
\begin{figure}
        \includegraphics[width=\columnwidth]{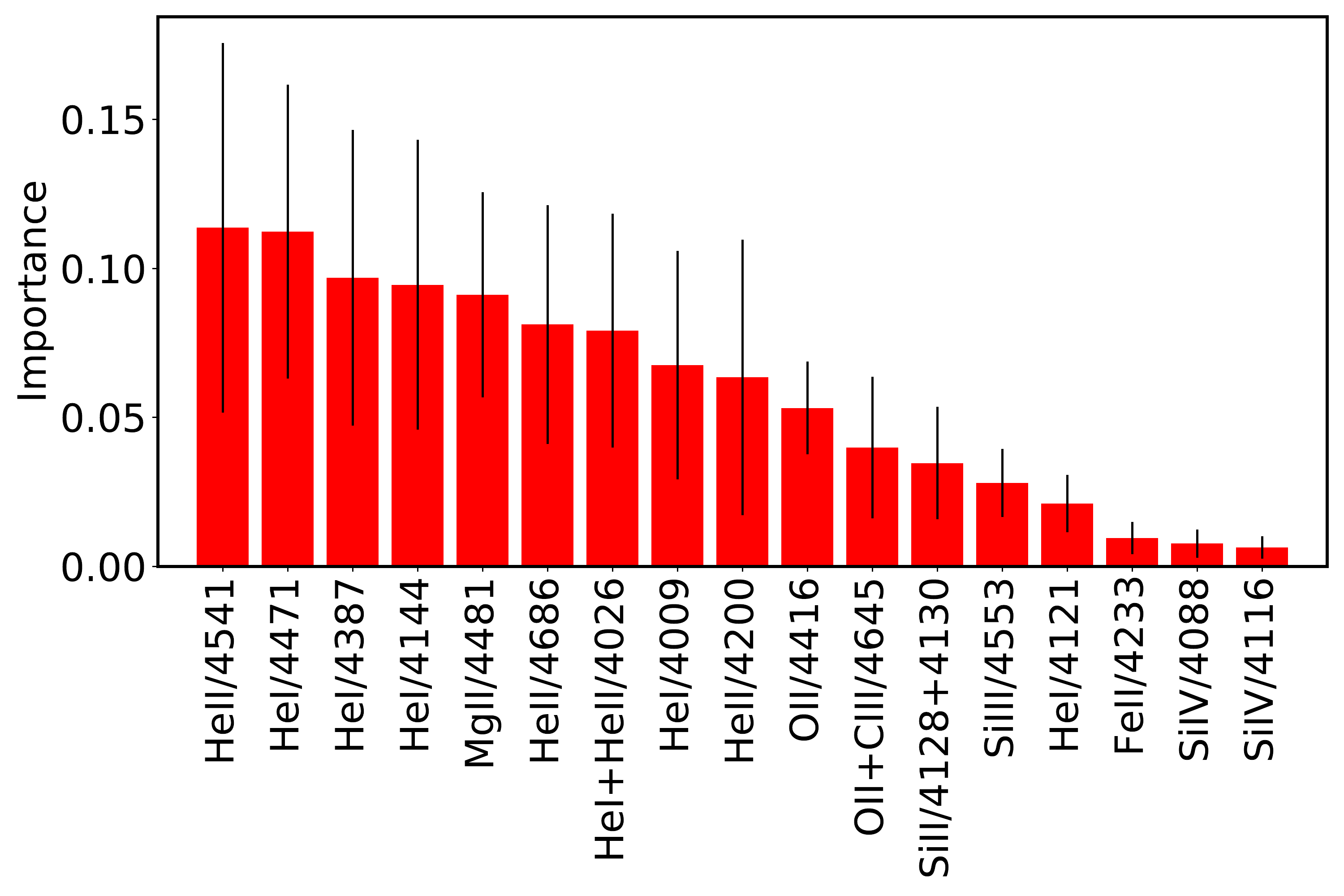}
        \caption{Feature of relative importance of spectral lines used in this work and its standard deviation.}
   \label{Fig.:features_importance}
\end{figure}

As a final test of  the credibility of the SFFS results, we compared them with a different method of feature selection that estimates how much each feature contributes during the training process and is included in \texttt{sklearn’s} RF implementation. Specifically, when a tree is trained, the decrease of each feature's Gini impurity can be computed. Afterwards, this impurity decrease can be averaged across all the trees in the forest, and the features are ranked according to the above measurement. The result of this feature importance quantification is shown in Fig. \ref{Fig.:features_importance}. The most important features are consistent with the spectral lines that have been selected by the SFFS algorithm until it reaches its plateau where any other addition of features does not significantly improve the performance of the algorithm. Furthermore, in both methods, the \ion{Fe}{II}/4233 $\mathbf{\AA}$, \ion{Si}{IV}/4088 $\mathbf{\AA}$, and \ion{Si}{IV}/4116 $\mathbf{\AA}$ lines appear to be insignificant for training the RF model; this result is also supported by their EWs distributions. In addition, the high-ranked spectral lines are in accordance with the SFFS results and consistent with the behavior of their EW distributions per spectral class. Overall, the results of the comparison between these two different techniques are consistent. Therefore, we are confident that the scheme presented in Table \ref{sec:features_table} is robust.\\

\section{Results}\label{sec:4}

\subsection{Evaluating the performance of the best RF and PRF models}\label{subsec:RF_PRF_best_models_evaluation}
After optimizing the algorithms and training the models, we evaluated their performance on the test set. For the evaluation of the algorithms, we used the confusion matrix, which is widely used in problems of statistical classification (e.g., \citealt{confusion_matrix,XRBs_classification}). The confusion matrix has a table layout that visualizes the performance of a classifier using the test subset. Each row of this table represents a true class, while each column represents a predicted class. Therefore, the confusion matrix shows the number of objects in each class versus the number of objects predicted by the model to belong to a particular class. In the ideal case, where the prediction ability of the model is 100\% correct, we would expect all objects to be on the diagonal of the matrix. 
For the interpretation of the confusion matrix and the assessment of the performance of a model, we used basic terms and metrics, which are defined in Table \ref{tab:metrics_table} (c.f., \citealt{Pearson}). Accuracy and Recall are the basic metrics that we use in this work since we are interested in how accurate our model is, as well as how correct the predicted spectral types are.

\begin{table*} 
    \centering
        \caption{Definition of the terms and the metrics that are used in this work for the valauation of the classifiers.}
            \begin{tabular}{lp{0.5\textwidth}cc}
            \hline
            Term & Definition & Formula \\
            \hline
            True Positive (TP) & An object is predicted from the classifier to belong to a spectral class, and it actually belongs to this class.&\\
            True Negative (TN) & An object is predicted from the classifier to not belong to a spectral class, and it actually does not belong to this class.& \\
            False Positive (FP) & An object is predicted from the classifier to belong to a spectral class, and it actually does not belong to this class. & \\
            False Negative (FN) &An object is predicted from the classifier to not belong to a spectral class, and it actually belongs to this class. & \\
            Accuracy& Number of objects that are predicted correctly from the classifier, over the total number of the tested sample & (TP+TN) / (TP+TN+FP+FN)\\
            Precision& Number of objects that are predicted correctly from the classifier to belong to a spectral class, over the total number of objects predicted by the classifier to belong to this class.& TP / (TP+FP)\\
            Recall& Number of objects that correctly predicted from the classifier to belong to a spectral class, over the total number of objects that actually belong to this class. &TP / (TP+FN)\\
            F1-score& The harmonic mean of the Recall and Precision. It is as an overall performance metric for the classifiers & 2TP / (2TP+FP+FN)\\
            \hline
            \end{tabular}
            \label{tab:metrics_table}
\end{table*}

\begin{figure*}
        \includegraphics[width=\columnwidth]{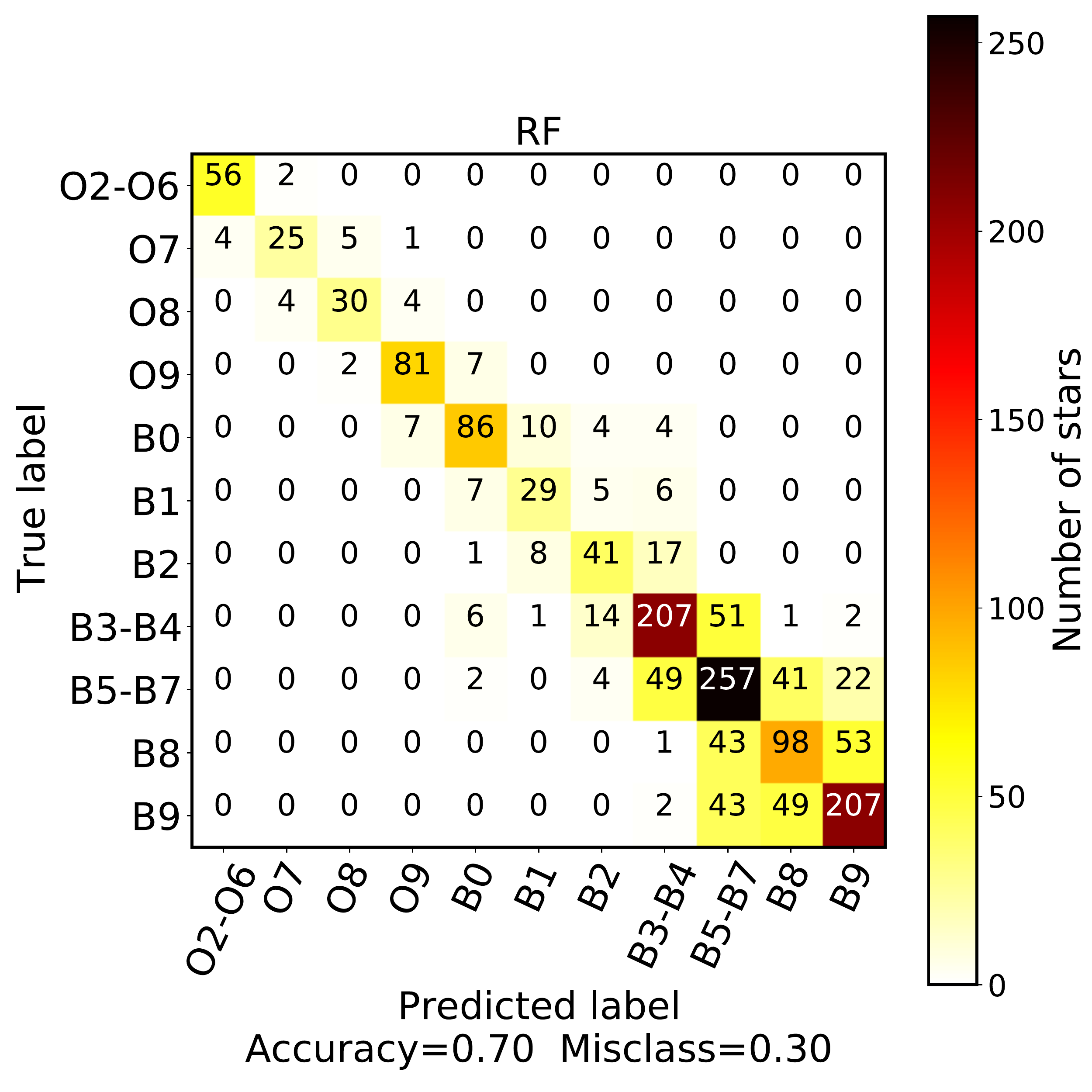}
        \includegraphics[width=\columnwidth]{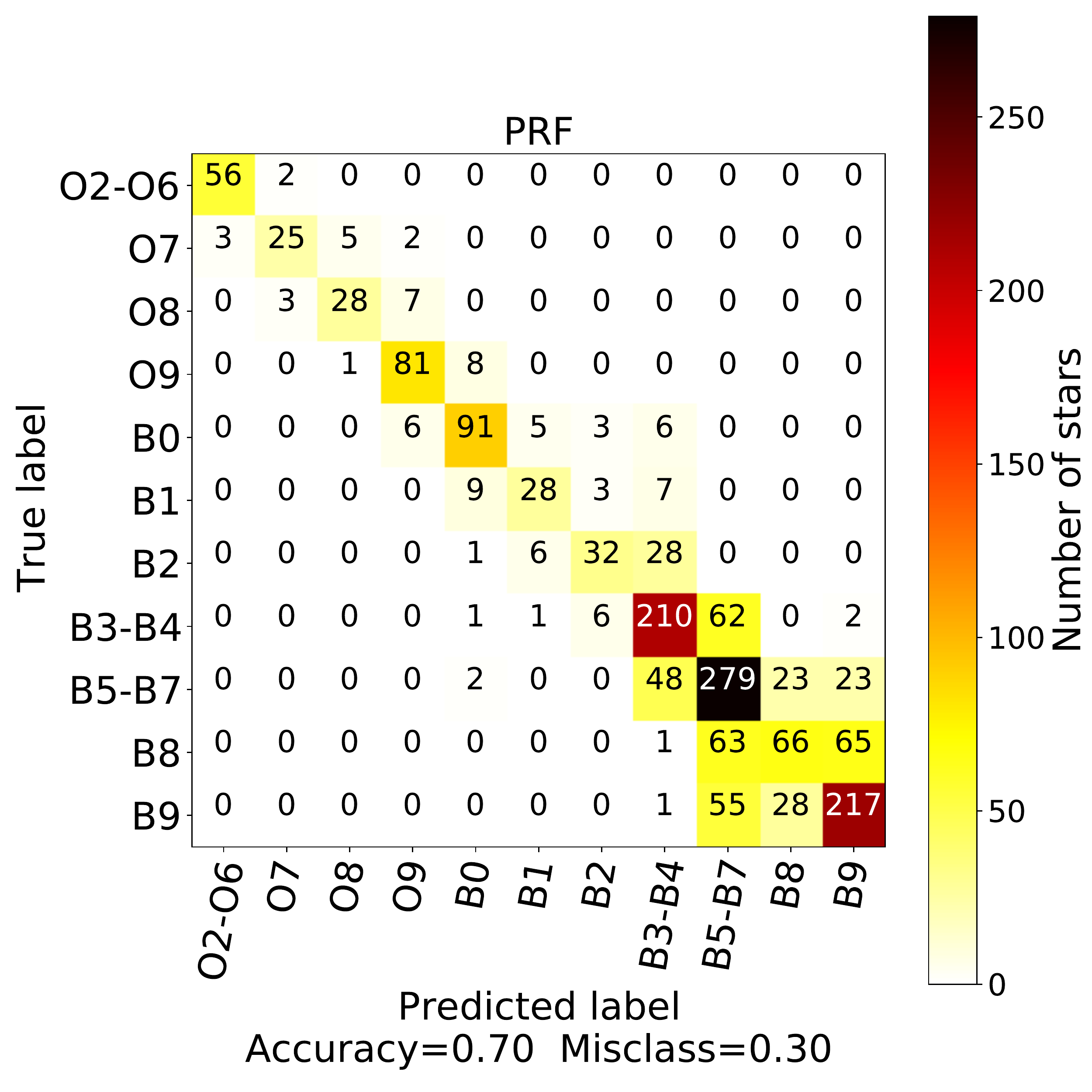}
        \begin{center}
        \includegraphics[width=\columnwidth]{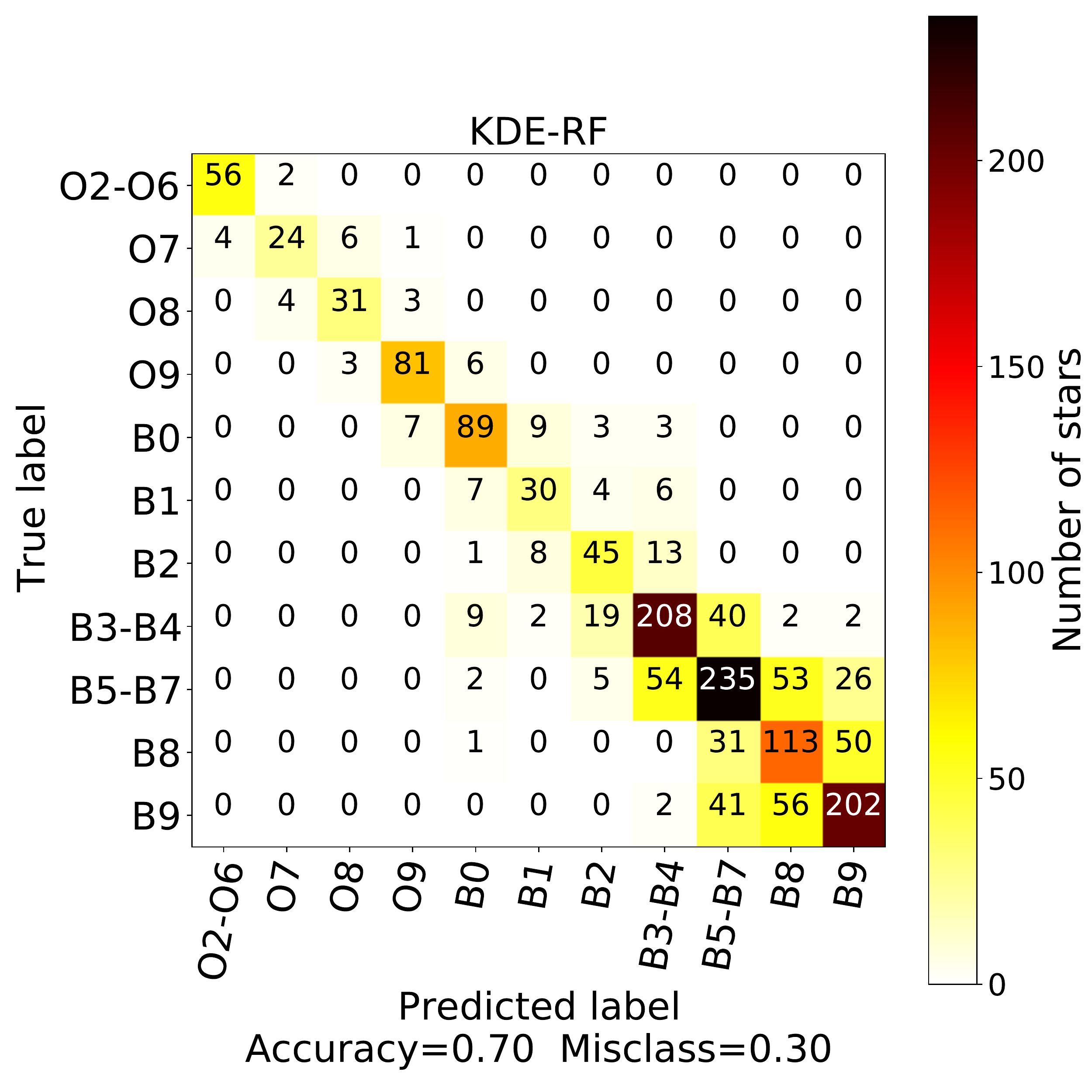}
        \end{center}
    \caption{Top left panel shows the confusion matrix of the best RF model applied to the test sample. The right panel shows the confusion matrix of the PRF best model applied to the same data set. The bottom panel shows the confusion matrix of the KDE-RF method. The overall accuracy is the same for all algorithms, $70$ $\%$, with the majority of misclassified objects belonging to neighboring classes, indicating the reliability of the algorithms.}
    \label{Fig.:Consusion matrix RF-PRF}
\end{figure*}

In the top left and top right panels of Fig. \ref{Fig.:Consusion matrix RF-PRF}, we present the confusion matrices of the RF and PRF models, respectively. Furthermore, in Table \ref{tab:best_models_metrics} we present the performance metrics per class for both algorithms, as well as the number of stars that were used as a test sample for each class. 

\begin{table*} 
    \centering
        \caption{Precision, recall, and F1-score metrics per spectral type as determined by the best models of RF, PRF, and our method applied to a test sample of 1599 stars.}
            \begin{tabular}{lccccccccccccc}
            \hline\hline
            Class & \multicolumn{3}{c}{RF} &
            \multicolumn{3}{c}{PRF}&\multicolumn{3}{c}{KDE-RF}&Test sample\\
            &Precision & Recall & F1-score &Precision & Recall & F1-score &Precision & Recall & F1-score\\
            \hline
            O2-O6&0.93&0.97&0.95&0.95&0.97&0.96&0.93&0.97&0.95&58\\
            O7&0.81&0.71&0.76&0.83&0.71&0.77&0.8&0.69&0.74&35\\
            O8&0.81&0.79&0.8&0.82&0.74&0.78&0.78&0.82&0.79&38\\
            O9&0.87&0.9&0.89&0.84&0.9&0.87&0.88&0.9&0.89&90\\
            B0&0.79&0.77&0.78&0.81&0.82&0.82&0.77&0.8&0.79&111\\
            B1&0.6&0.62&0.61&0.7&0.6&0.64&0.61&0.64&0.62&47\\
            B2&0.6&0.61&0.61&0.73&0.48&0.58&0.59&0.67&0.63&67\\
            B3-B4&0.72&0.73&0.73&0.7&0.74&0.72&0.73&0.74&0.73&282\\
            B5-B7&0.65&0.69&0.67&0.61&0.74&0.67&0.68&0.63&0.65&375\\
            B8&0.52&0.5&0.51&0.56&0.34&0.42&0.5&0.58&0.54&195\\
            B9&0.73&0.69&0.71&0.71&0.72&0.71&0.72&0.67&0.7&301\\
            \hline
            \end{tabular}
            \label{tab:best_models_metrics}
\end{table*}

As can be seen from the confusion matrices, both the RF and the PRF models have an overall accuracy of $\sim 70 \%$. This means that their performance is similar, although their training method is different in terms of including measurement uncertainties. In addition, the k-fold validation for both models gives a CV score of $69 \pm 2\%$ for RF and $70.0 \pm 0.5\%$ for PRF. To further assess the uncertainty of our classifiers we repeated the k-fold CV test three times with a different random split of the data set in k-folds at each time. Afterwards, we calculated the average accuracy and the standard deviation across all folds and all repeats. This repeated k-fold returns effectively the same scores ($69 \pm 1\%$ and $70 \pm 1\%$, respectively).
These scores are high if we consider the complexity of such multi-class classification problems and the fact that for the training we used a diverse sample including data from a variety of surveys with different observational setups, which also suffer from non-negligible measurement errors for the spectral features. In addition, the labels used for the training also include an intrinsic uncertainty. The fact that the majority of misclassified stars belongs to neighboring classes, combined with the low standard deviation of the CV scores, indicates the reliability and the stability of both algorithms. 

A careful examination of the metrics for both algorithms (Table \ref{tab:best_models_metrics}), reveals that their behavior is similar not only in terms of overall score but also with respect to the scores per class. More specifically, the two models show relatively poor performance in the case of early B-type stars. In contrast, they perform better when they classify spectral types of O and late B-type stars. Since we are interested in the true positives, we focused on the recall metric. Comparing the values of the same classes in both models we see that the highest scores are for class O2-O6 (0.97 for both RF and PRF) and the lowest scores are for class B8 (0.50 and 0.34, respectively). In addition, high scores are reached for the O7, O8, O9, and B0, as well as for B3-B4 and B9, while for the remaining classes (e.g.,  B1, B2, B5-B7) the scores are lower, indicating a difficulty for both models to distinguish these spectral classes.

\subsection{Development of an independent method for incorporating uncertainties}\label{subsec:Simulated_data}
The regular RF model does not handle measurement uncertainties on the features during the training process, contrary to the PRF. 
Furthermore, both methods are sensitive to the sampling of the feature distributions for each class. In our case, this is a problem since many classes are significantly underrepresented. While the imbalance between the different classes can be accounted for with the \texttt{class\_weight='balanced\_subsample'} flag, this cannot account for the scarce sampling of the feature parameter space in the poorly represented classes.   
With this in mind, we devised an alternative approach to the standard data augmentation methods where we incorporate in the sampling process a kernel density estimation (KDE). 

A KDE is a nonparametric way to estimate the probability density function (PDF) of a data set without any previous knowledge of its distribution. More specifically, the KDE uses a mixture consisting of a predefined kernel component per data point. The sum of the components for all the data points essentially results in a nonparametric estimator of the density of the overall sample. In our case, we adopted a Gaussian kernel.  

Following our previous training/test sample split (70/30 \%), we first defined the training sample. Then, based on this sample we created a multidimensional histogram per spectral class (11 in total), which includes the EW for the spectral lines in the considered features (see Table \ref{sec:features_table}).

Essentially, each subset is a different seventeen-dimensional dataset encapsulating the joint distribution of the EW of the characteristic spectral lines based on the measured EW of the objects of each class. Then, we applied a separate Gaussian KDE per subset (i.e., class) and we determined 11 different seventeen-dimensional PDFs, providing the joint distribution of the EW for all features of interest per class. The bandwidth (smoothing parameter) of the applied kernels was chosen to be equal to the medians of the uncertainties of the EW values for all the features in each class. Once we had calculated the PDFs, we drew 1000 sets of EWs per class. This way, we created a new balanced data set of 11000 mock data representing objects very similar to the real ones, making it the proper training data set for the RF algorithm. This approach has the following advantages: (a) it smooths the sparsely sampled feature distributions of the underrepresented classes; (b) it accounts for the correlated behavior of the different features within each class and between classes (which could increase the efficiency of the model); and (c) it resolves the bias resulting from the significant imbalance of the training data for different classes. 

For the training, we used the same optimized RF configuration (see Table \ref{tab:opt_hyperpar}), since the data distribution on which the model was optimized is the parent distribution of the artificial data set, and consequently we do not expect significant differences in the values of the hyperparameters. As a final step, we tested the performance of our method on the same test set that we used for the evaluation of the RF and PRF. In the bottom panel of Figure \ref{Fig.:Consusion matrix RF-PRF}, we present the final confusion matrix of our method, while the performance metrics per class are tabulated in Table \ref{tab:best_models_metrics}. This method will be referred to as the KDE-RF.

As we see, the overall accuracy of our method is $\sim 70 \%,$ and the corresponding k-fold CV score is $69.0 \pm 0.5 \%$. In addition, the repeated k-fold gives a score of $68 \pm 1 \%$. Furthermore, in terms of the performance of the model for the individual spectral classes, we see again that the classifier better predicts the early-O-type and late-B-type stars in comparison to the early B types. The highest and lowest recall score is for the O2-O6 (0.97) and for B8 (0.60) classes, respectively. In general, our method demonstrates the same performance within the errors as the other methods we discussed in Section \ref{subsec:RF_PRF_best_models_evaluation}, but recall is slightly better in the underperforming classes (e.g., B8). This similarity of the performance for the three different approaches once again indicates the stability and the reliability of the methods. 

In order to test to what degree the performance of the model is affected by the correlation between the EW of the spectral lines for each star, we performed the following Monte Carlo analysis. For each star, we drew 100 samples of EW values per line from Gaussian distributions with means and standard deviations equal to those derived from the spectral fit analysis. This way, we preserved the correlation between the strength of the different spectral lines for each star belonging to each spectral class we considered. We then performed the RF analysis using this sample of 373000 simulated objects, which gave an overall accuracy of 69\% and a confusion matrix virtually identical to those presented in Figure \ref{Fig.:Consusion matrix RF-PRF}. This indicates that the dominant factor determining the behavior of the RF classifier is the intrinsic scatter (spread) of the EWs for each spectral class.\\

\section{Discussion} \label{sec:5}

\subsection{Summary of the results }\label{subsec:comments_on_the_reults}

\begin{figure}
        \includegraphics[width=\columnwidth]{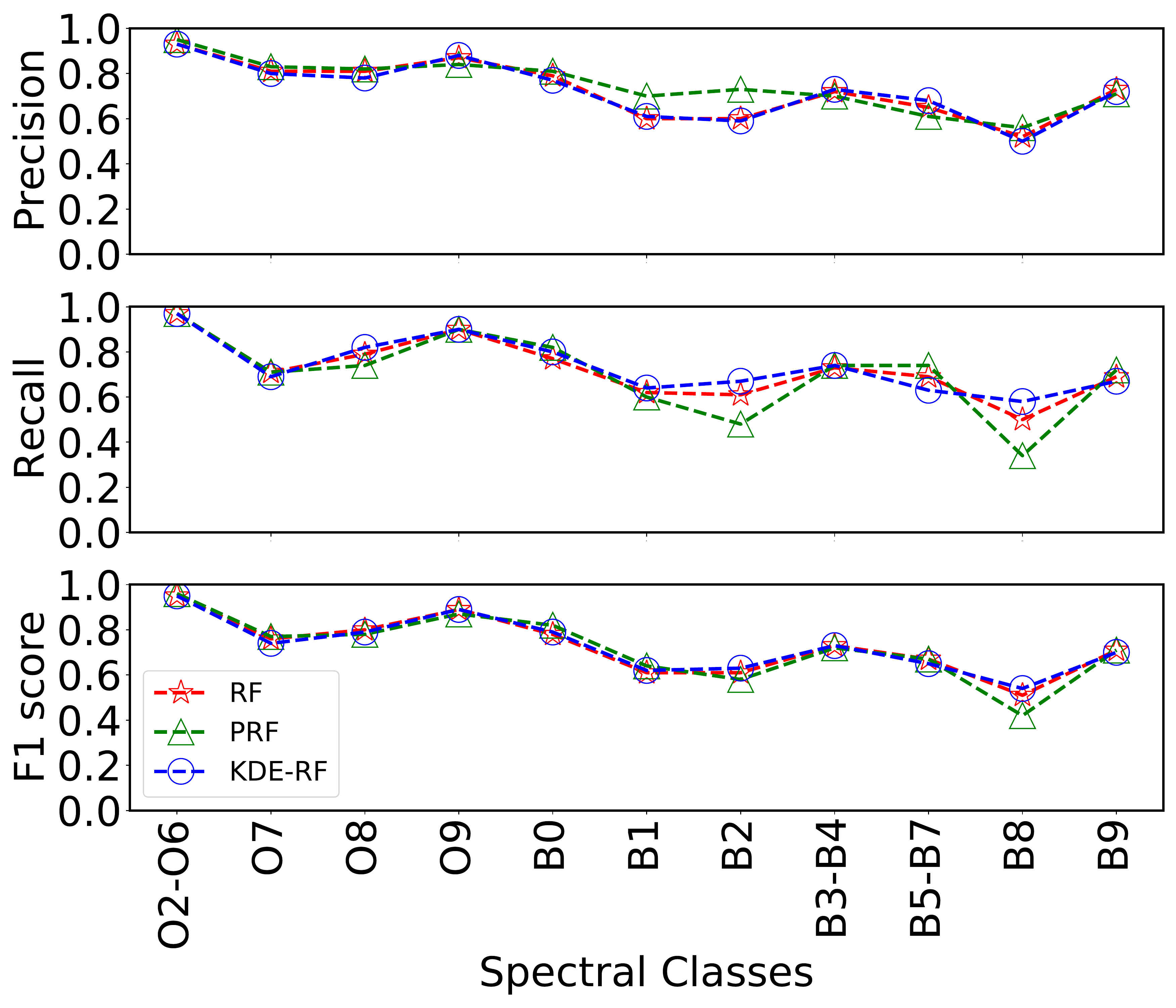}
        \caption{Comparison of the three models used in this work. Stars and triangles correspond to the results from the RF and PRF algorithms, respectively. The results from our method are shown with open circles. We find that all the methods are consistent, indicating the robustness of the RF algorithm for the automated spectral classification of OB stars. The majority of spectral classes are predicted with a score higher than 65$\%$, while the lower performance of some classes is either due to their inadequate representation in the initial sample or due to overlap of the EW distributions of their characteristic spectral lines with those of other classes (see Section \ref{subsec:comments_on_the_reults}). }
   \label{Fig.:metrics_comparison_3_models}
\end{figure}

We used three different RF-based methods to assess if this algorithm can be used for the spectral classification of OB-type stars into their subtypes. We find that all the models have similar performance, both in their overall score, and also with respect to the individual spectral classes. Moreover, the fact that for almost all the classes the misclassified objects belong to neighboring classes indicates the success of our classifier. With such classification problems, the classes are contiguous and ordered based on the continuous nature of the underlying physical parameter, which in the case of spectral types is the temperature. For the direct comparison of the three methods, in Fig. \ref{Fig.:metrics_comparison_3_models} we present how their basic metrics vary per spectral class. Essentially, we plot the values given in Table \ref{tab:best_models_metrics}, and as we can see, the precision, recall, and F1-score exhibit almost the same variation with respect to the different classes for all models. Focusing on the F1-score, which is an average estimation for both precision and recall, we see that the best performing classes are O2-O6, O7, O8, O9, B0, B3-B4, B5-B7, and B9 with a score $>0.65$, while the B1, B2, and B8 classes reach scores close to $0.60$, $0.40$, and $0.5,$ respectively. This means that all models are able to classify the majority of the spectral classes with relatively high accuracy, indicating that the RF algorithm can be used for such multiclass classification problems.

The reason behind the lower F1 score for these classes can be seen by examining the EW distributions (Figure \ref{Fig.:ew_distribs}) of the specific spectral lines that distinguish each class from its neighboring classes. In particular, according to Table 2 from \cite{Maravelias}, the main spectral lines that help to  discriminate between stars of spectral classes B1 and B2 are \ion{Si}{IV}/4088,\ion{Si}{IV}/4116 and \ion{Si}{III}/4553 $\mathbf{\AA}$. These lines show a strong overlap of their EW values, not only for the B1 and B2 spectral classes, but also for all the examined classes. That means that the RF classifier cannot learn anything from these features resulting in poor performance for these spectral classes. Exactly the same factor is responsible for the poor performance of the B8 class, although in this case the characteristic spectral lines that characterize it are \ion{He}{I}/4144 and \ion{Si}{II}/4128+4130. As is shown from their EW distributions, there is no clear separation of the B8 class from its adjacent classes B5-B7 and B9. The lack of useful information from these spectral lines, resulting from the strong overlap  in all spectral classes, is also reflected in Figure \ref{Fig.:features_importance}, where we see that the \ion{Si}{IV}/4088-4116 and \ion{Si}{II}/4128+4130 lines are among the less diagnostically important. Although \ion{He}{I}/4144 does not help to distinguish between the B8 and B9 classes, we see that its importance is higher due to the clear separation of the EW values for earlier spectral classes. 

\begin{figure}
        \includegraphics[width=\columnwidth]{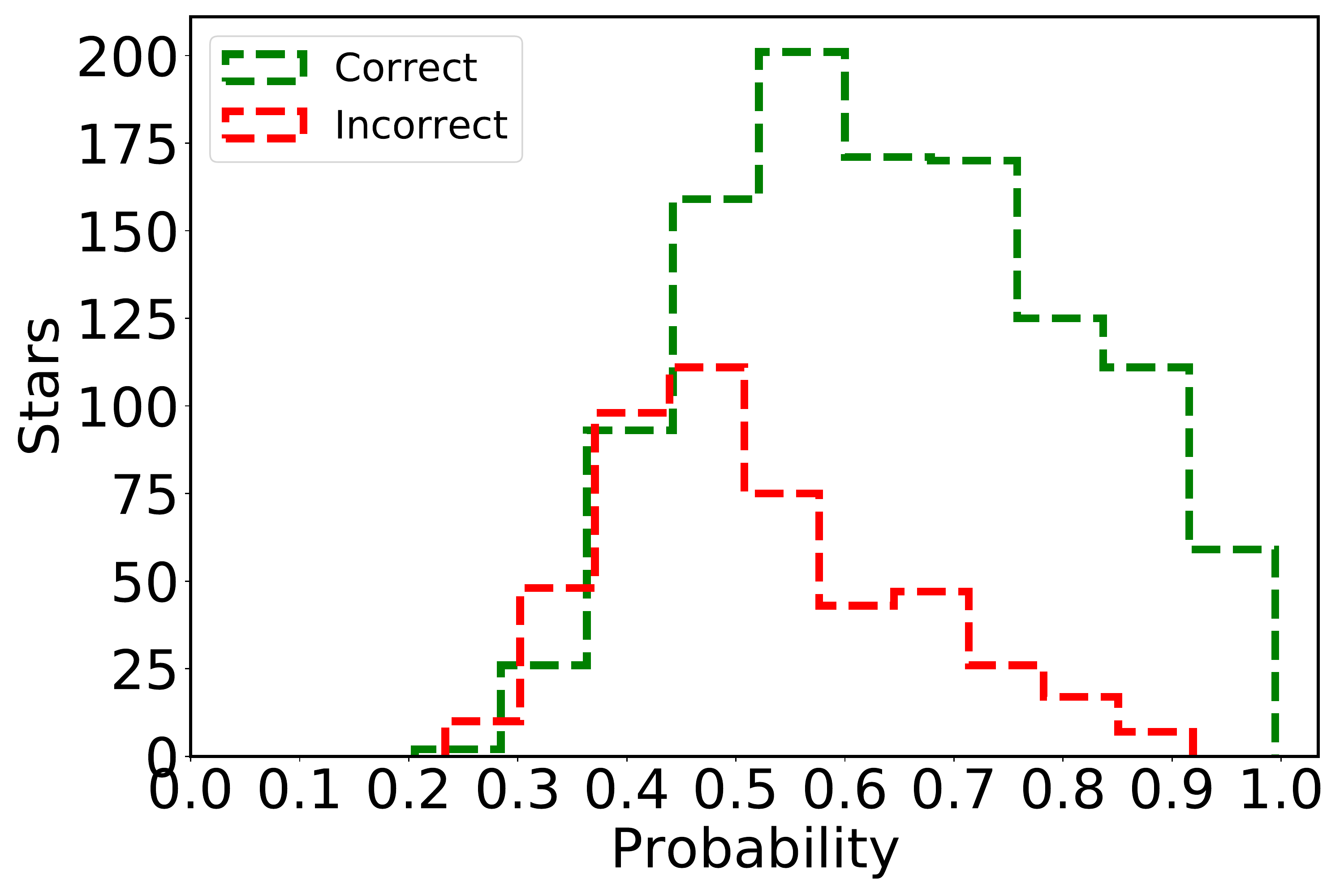}
        \caption{Probability distributions between the sources that were predicted correctly (green dashed line) and incorrectly (red dashed line). The medians of the two distributions are 0.64 and 0.49, respectively. Based on these distributions, we define three flags regarding the reliability of our results. In particular, suggested spectral types with probability > 0.64 are flagged as strong candidates, while those with probability < 0.48 are flagged as candidates. For the intermediate range, 0.48-0.64, they are flagged as good candidates.}
   \label{Fig.:correct_incorrect_probs}
\end{figure}

The predicted class of the RF algorithm is the one with the highest probability in the returned probability distribution. In Fig. \ref{Fig.:correct_incorrect_probs}, we plot the number of test sources predicted correctly and incorrectly versus the probability inferred from the RF algorithm. The two distributions have medians of 0.64 and 0.49, respectively, indicating that, in general, correctly predicted objects have higher probabilities than the incorrectly predicted ones. Specifically, above the threshold of 0.64, we have 558 true positives and 102 false negatives which means that the accuracy is $\sim 85\%$ in the reliable classification regime. In contrast, in the lower reliability probability range 0.49-0.64, the true positives are 334 and the false negatives are 139, corresponding to an accuracy of $\sim 71\%$. Below the threshold of 0.49, we have 224 true positives and 241 false negatives, or $\sim 48\%$ accuracy. With this in mind, we define three different confidence levels for the predicted spectral classes based on the aforementioned probability distributions: a) above 0.64 (strong candidate); b) 0.48-0.64 (good candidate); c) below 0.48 (candidate). 

\subsection{Sensitivity of classification on measurement and label uncertainty}\label{subsec:error_sensitivity}

In our analysis, we used the PRF algorithm for the first time (to our knowledge) in order to asses its performance in real observational data and compare it with the regular RF. The PRF  was developed to incorporate the feature uncertainties during the training stage, and it is expected to exhibit an improvement of the order of 10$\%$ in accuracy compared to RF \citep{PRF}. However, our results show that the PRF and KDE-RF do not provide an improvement over the RF. The reason is that the performance of the algorithms is not driven by statistical scatter due to measurement errors. Instead, is mainly driven by the intrinsic scatter of the EW values within the range of the examined spectral classes, resulting in strong overlap between the EW distributions. This can also be seen from the Fig. \ref{Fig.:ew_fit_band_comparison}, where the median measurement uncertainties per spectral class are significantly lower than the scatter of the measurements. Furthermore, working on a real data set of observational data (spectra) rather than a synthetic one (as used in the development of the PRF method; \citealt{PRF}), may result in systematic errors due to noise, which corresponds to the use of different detectors and surveys.  

Besides the measurement uncertainties, in this type of classification problems there are also uncertainties in the labels. The visual inspection of a stellar spectrum and the estimation of its spectral type is a relatively subjective process. It is very common for different people examining the same spectrum to conclude on slightly different spectral types. This label uncertainty exists even in spectral types that are assigned to stars through a template comparison or other automated procedure. Although with the PRF algorithm these label uncertainties can be included in the training stage, we did not apply it in this work because they are not provided in most of the spectroscopic surveys we used. Only the LAMOST survey \cite{OB_catalog_lamost} includes this information, claiming that all the spectra that were classified with the MKCLASS software package \citep{MKCLASS} are correct, with one subtype uncertainty. Because there is no easy way to work around this problem and the rejection of the other three surveys would limit our sample (especially with earlier type stars or lower metallicites; i.e., for SMC and LMC), we opted not to implement this option. Nonetheless, our results as illustrated in the confusion matrices (Fig. \ref{Fig.:Consusion matrix RF-PRF}) show that the classification uncertainty is $\sim \pm 1$ class for later spectral classes (i.e., B5-B7,B8,B9) or better.  

\subsection{Volume of simulated training data}
\label{subsec:training_simulated_data_volume}
During the development of the KDE-RF method for the incorporation of uncertainties, we trained our model using a new balanced data set of mock data (see Section \ref{subsec:Simulated_data}). In particular, we chose to randomly produce feature values for 1000 stars per spectral type, thus creating an overall training sample of 11000 simulated data. However, an interesting question that arises is how the size of the sample affects the model's performance.

\begin{figure}
        \includegraphics[width=\columnwidth]{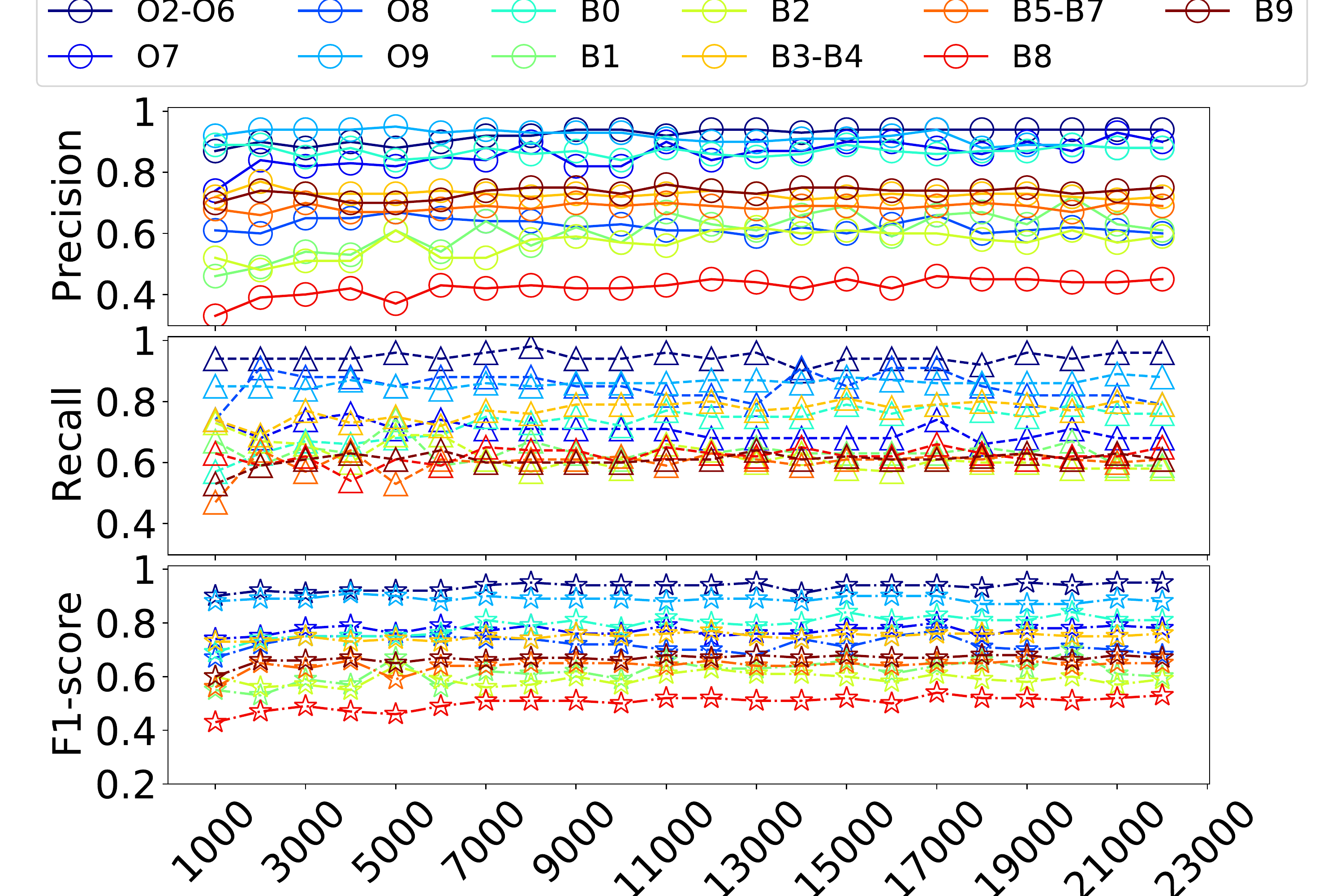}
        \caption{Precision, recall, and F1 score as a function of the volume of the total simulated data used for the training of our RF-based model developed in Section \ref{subsec:Simulated_data}. The total number of mock data divided by the total number of classes gives the number of mock data per class. The model's performance does not depend on the size of the training sample, and it is mainly constrained from the parent distribution of the real data. }
   \label{Fig.:training_data_volume}
\end{figure}

To address this question, we followed a similar method to \cite{Clarke} by adjusting the mock training sample and assessing the model's performance. We ranged the sample size from 100 to 2000 objects in steps of 100 objects for each of the 11 classes. 
In Fig. \ref{Fig.:training_data_volume}, we present how the basic metrics per class vary as a function of the number of the produced simulated data. The precision, recall, and F1 score are almost constant for all the spectral classes, indicating that the performance of our model does not depend on the volume of the mock data used for its training. Based on the above metrics, we see that even a ${\sim}$1000 strong sample (i.e., ${\sim}$100 per class) provides adequate sampling of the parent joint distribution of EW (as determined from the 17-D KDE). This could be the result of intrinsic scatter (or uncertainties) that smear the distribution and hide any complexity, in combination with smooth variations in the line strength as the temperature of a stellar atmosphere gradually decreases toward later spectral types. The small variations in the precision, recall, and F1-score values are due to the inherent stochasticity in the training procedure of the RF algorithm (random selection of features and samples for the construction of each decision tree in the RF). For the same reason, the results will not be identical in each realization of the algorithm, but they will definitely be close to the average score value that the best model can reach.

\subsection{Application of the RF best model on different science cases}\label{subsec:application}
The ultimate goal of this work is the application of the developed method on spectroscopic data sets of candidate OB-type stars for fast and reliable spectral classification. For this reason, we applied our best RF configuration to a number of science cases involving data of different qualities, providing examples of potential usage. During the data preparation for the analysis, we measured their EWs (following the same procedures described in Section \ref{subsec:EWs_EWs_errors}).

For the first application, we used data from the IACOB project \citep{IACOB_2011_a,IACOB_2011_b,IACOB_2015}. In particular, we used spectra\footnote{The wavelength range of the observations is $\sim 3700-7000$ $\AA$ and the associated resolving power R of the spectra is 25000/46000/65000 and 85000 for FIES and HERMES spectrographs, respectively.} from DR1 and DR2, which includes high-resolution and good S/N ($\geq$ 200) spectra from 328 Galactic massive OB-type stars with luminosity classes I-V. Although the IACOB spectra are of high quality, their spectral types are not well-determined since they come from different sources. With this in mind, we decided to revisit the spectral types of the IACOB project and suggest spectral-type classification based on the consistent application of our method. In Table \ref{tab:Application_results}, we present the results, and for comparison we also tabulate the spectral types provided from IACOB. We find that for 215 out of 328 stars, the suggested spectral classes from our model agree with those reported in the IACOB database. This score of $\sim 66 \%$ is very close to the maximum accuracy that our classifier can reach. Furthermore, for the remaining 113 stars, the algorithm predicts neighboring spectral classes. In addition, regarding the confidence level of the predictions, we find that $\sim 42 \%$ of the suggested spectral classes are flagged as "strong candidates", $\sim 30 \%$ as "good candidates", and only $\sim 28 \%$ are flagged as "candidates". The results are very encouraging since they confirm that our model works properly on totally blind data sets, indicating the robustness and the reliability of our classifier.

Our second application is focused on Galactic BeXRBs. We used 12 spectra of the donor stars for nine well-studied BeXRB systems obtained with the 1.3 m teleskope at the Skinakas Observatory \footnote{The wavelength range of the observations is $\sim 3616-5717$ $\AA$ and the resolving power is $R \sim$ 2500.}. As is shown in Table \ref{tab:Application_results}, our model predicts spectral classes very close to those derived from manual classification. Notably, if we take into account vague uncertainties in our model, we can say with safety that the RF model works adequately. In addition, for each of the systems, 4U2206+54, IGR J06074+2205, and 1A 0535+262 were observed in two instances, and thus we tested our model on spectra of different qualities. For IGR J06074+2205 and 4U2206+54, we obtain the same correct prediction from both sources, while for 1A 0535+262 we obtain incorrect predictions from both observations of the source. However, these misclassifications are due to poor measurements, even for the most characteristic spectral lines of the B0 spectral class (e.g., \ion{He}{II}/4541,\ion{He}{II}/4686 etc): almost all the spectral lines were replaced with magic numbers, indicating the low quality of these spectra. To conclude, our classifier performs well even in the special class of Oe and Be stars.

The above applications of the RF model prove that our proposed spectral-class classifier works well on various instances of real-world observational data constrained only by the quality range of the data on which it was trained. For this reason, we suggest to potential users of our tool that its performance is optimal for data with  S/N $>50,$ and it requires spectral line measurements in the $\sim 3900-4900$ $\mathbf{\AA}$ wavelength range and $\sim 1500-4000$
resolution.
\begin{landscape}
    \begin{table}
        \begin{threeparttable}

        \caption{Final products of our model's applications.}
        \begin{tabular}{ccccccccccccccccc}
        \hline\hline\\
        (1)&(2)&(3)&(4)&(5)& & & & & &(6)&\\
        IACOB ID & Publ. & RF & Prob.&Confidence level& \multicolumn{11}{c}{Probability per class}\\ 
        & & & & &B0&B1&B2&B3-B4&B5-B7&B8&B9&O2-O6&O7&O8&O9\\
        \hline
HD1279&B8&B5-B7&0.419&candidate&0.014&0.0&0.001&0.029&0.419&0.316&0.193&0.006&0.007&0.009&0.005\\
HD1337&O9&O9&0.9036&strong candidate&0.044&0.001&0.0&0.001&0.001&0.0&0.0&0.0&0.008&0.042&0.904\\
HD1438&B8&B8&0.4052&candidate&0.002&0.0&0.001&0.011&0.255&0.405&0.325&0.0&0.0&0.0&0.001\\
HD1976&B5&B3-B4&0.8691&strong candidate&0.001&0.006&0.034&0.869&0.084&0.001&0.005&0.0&0.0&0.0&0.0\\
HD2626&B7&B5-B7&0.6808&strong candidate&0.002&0.001&0.01&0.107&0.681&0.098&0.101&0.0&0.0&0.0&0.0\\
HD2905&B0&B1&0.4515&candidate&0.266&0.451&0.039&0.08&0.047&0.002&0.01&0.0&0.003&0.004&0.096\\
HD3240&B7&B5-B7&0.3528&candidate&0.002&0.0&0.006&0.03&0.353&0.342&0.263&0.0&0.002&0.001&0.001\\
HD3369&B5&B3-B4&0.7726&strong candidate&0.015&0.017&0.065&0.773&0.117&0.0&0.011&0.0&0.0&0.0&0.002\\
HD4142&B5&B3-B4&0.7976&strong candidate&0.007&0.007&0.045&0.798&0.128&0.002&0.011&0.0&0.0&0.0&0.001\\
HD4382&B8&B5-B7&0.3811&candidate&0.001&0.002&0.004&0.015&0.381&0.347&0.251&0.0&0.0&0.0&0.0\\
HD4636&B9&B5-B7&0.4275&candidate&0.004&0.001&0.0&0.017&0.427&0.337&0.189&0.005&0.006&0.005&0.009\\
HD4727&B5&B3-B4&0.3826&candidate&0.191&0.114&0.194&0.383&0.094&0.004&0.011&0.0&0.001&0.001&0.007\\
HD14272&B8&B5-B7&0.3388&candidate&0.001&0.0&0.005&0.027&0.339&0.318&0.31&0.0&0.0&0.0&0.0\\
HD14818&B2&B1&0.3782&candidate&0.057&0.378&0.108&0.321&0.109&0.005&0.015&0.0&0.001&0.001&0.004\\
HD14947&O4&O2-O6&0.9622&strong candidate&0.001&0.0&0.0&0.0&0.001&0.001&0.002&0.962&0.021&0.002&0.01\\
HD15137&O9&O9&0.9234&strong candidate&0.065&0.001&0.0&0.003&0.001&0.0&0.0&0.0&0.0&0.007&0.923\\
HD15318&B9&B9&0.8522&strong candidate&0.001&0.003&0.0&0.005&0.037&0.101&0.852&0.001&0.0&0.0&0.0\\
HD15558&O4&O2-O6&0.8823&strong candidate&0.005&0.002&0.0&0.0&0.001&0.001&0.002&0.882&0.038&0.003&0.065\\
HD15570&O4&O2-O6&0.876&strong candidate&0.004&0.001&0.001&0.001&0.003&0.011&0.007&0.876&0.053&0.012&0.031\\
HD16046&B9&B9&0.8961&strong candidate&0.0&0.002&0.0&0.005&0.024&0.071&0.896&0.002&0.0&0.0&0.0\\
        \hline\\        
 HMXBs ID &Publ. & RF & Prob.&Confidence level& \multicolumn{11}{c}{Probability per class}\\ 
            & & & & &B0&B1&B2&B3-B4&B5-B7&B8&B9&O2-O6&O7&O8&O9\\
            \hline
            
1A 0535+262\_1st&B0&O2-O6&0.3274&candidate&0.107&0.009&0.001&0.013&0.023&0.021&0.02&0.327&0.243&0.113&0.123\\
1A 0535+262\_2nd&B0&B5-B7&0.2817&candidate&0.215&0.068&0.006&0.148&0.282&0.09&0.094&0.012&0.028&0.015&0.042\\
SAX J2103.5+4545&B0&O9&0.4197&candidate&0.185&0.029&0.009&0.056&0.078&0.012&0.02&0.003&0.053&0.136&0.42\\
RX J0440.9+4431\_1st&B0&B0&0.3124&candidate&0.312&0.169&0.031&0.185&0.118&0.034&0.041&0.015&0.019&0.023&0.052\\
RX J0440.9+4431\_2nd&B0&B0&0.2393&candidate&0.239&0.044&0.001&0.03&0.064&0.043&0.044&0.08&0.103&0.143&0.207\\
IGR J06074+2205\_1st&B0&B0&0.3432&candidate&0.343&0.245&0.045&0.127&0.085&0.048&0.035&0.012&0.014&0.009&0.038\\
IGR J06074+2205\_2nd&B0&B0&0.3229&candidate&0.323&0.281&0.159&0.203&0.018&0.003&0.002&0.0&0.002&0.002&0.007\\
RX J0240.4+6112&B0&O9&0.63067&good candidate&0.277&0.022&0.008&0.009&0.006&0.001&0.001&0.0&0.008&0.039&0.631\\
2S0114+650&B1&B1&0.2458&candidate&0.222&0.246&0.023&0.086&0.125&0.053&0.081&0.03&0.024&0.035&0.077\\
IGR J21343+473&B1&B0&0.3145&candidate&0.314&0.239&0.309&0.106&0.012&0.002&0.002&0.0&0.0&0.0&0.015\\
RX J0146.9+6121&B1&B0&0.3594&candidate&0.359&0.093&0.004&0.035&0.062&0.03&0.053&0.018&0.016&0.035&0.295\\
4U2206+54\_1st&O9&O9&0.3997&candidate&0.231&0.025&0.002&0.016&0.023&0.011&0.02&0.067&0.072&0.134&0.4\\
4U2206+54\_2nd&O9&O9&0.497&candidate&0.258&0.011&0.002&0.016&0.007&0.002&0.002&0.011&0.074&0.121&0.497\\         
            \hline\\
            \end{tabular}
            \label{tab:Application_results}
            \begin{tablenotes}
                \item Note 1: Only 20 objects from IACOB survey are tabulated. The full version will be available in electronic form at the CDS via anonymous ftp to cdsarc.u-strasbg.fr (130.79.128.5) or via  http://cdsweb.u-strasbg.fr/cgi-bin/qcat?J/A+A/.
                \item Note 2: Columns description: (1) Object's ID, (2) Published spectral type, (3) Predicted spectral class, (4) Probability of predicted spectral class, (5) Confidence level flag by following the scheme described at Section \ref{subsec:comments_on_the_reults}, (6) Probability per class.
                \item Note 3: In the case of HMXBs spectra, 1st and 2nd indicate multiple spectra of the same object considered in our analysis.
\end{tablenotes}
  \end{threeparttable}
  
    \end{table}
\end{landscape}

\section{Conclusions}\label{sec:6}
The main results from this work can be summarized as follows:
\begin{enumerate}
    \item 
    We developed a new automated tool for the spectral classification of OB stars into their sub-classes, which is agnostic to the luminosity class of a star, and as much as possible, trained in a wide range of metallicities. The well-known RF algorithm, which is closer to the traditional way of optical spectroscopy in comparison with other automated methods is at the core of this new tool. In addition, we incorporated the feature uncertainties in our analysis by using the new PRF algorithm, as well as by developing our KDE-RF method based on data simulation. This method is based on optical spectra of O-B type stars in the $\sim 3900-4900$ $\AA$ wavelength range and $> \sim 50$ S/N range.
    \item
    Comparing the three models, we find a similar accuracy score of $\sim 70\%.$  This score is high if we consider the complexity of such multiclass classification problems (i.e., 11 classes), the intrinsic scatter of the EW distributions within the examined spectral classes and the diversity of the training set (resulting from the use data obtained from different surveys with different observing strategies). Given the lack of differences in the scores of the three methods, no method is considered superior. However, the proposed KDE-RF method is helpful for the data augmentation of sparsely populated samples such as those for the early-type stars. The similarity in the performance of our models indicates the robustness and the reliability of the RF algorithm when it is used for spectral classification of early-type stars that are isolated or companions in BeXRBs.  
    \item Using two different feature selection methods alongside the EW distribution per spectral line of our sample, we find that the full set of 17 spectral lines is needed to reach the maximum performance per spectral class. However, in Appendix \ref{10-feature scheme} we propose a reduced ten-features scheme that can be applied to large data sets with lower S/Ns ($\sim 20-50$) and where the measurement of weaker spectral lines is not reliable.
    \item In terms of the F1-score, which is an average estimation for both precision and recall, we see that the best performing classes are the O2-O6, O7,O8, O9, B0, B3-B4, B5-B7, and B9, with a score $> 0.65$, while the B1, B2, and B8 classes reach scores close to 0.60, 0.40, and 0.5, respectively. This means that all models are able to classify, with relatively high accuracy, the majority of the spectral classes, indicating that the RF algorithm can be used for such multiclass classification problems.
    \item We used the PRF algorithm for the first time and find that although it is more computationally intensive, it performs well when it is applied directly to real-world observational data. The general accuracy score does not change in our case. The uncertainty in the values of the features does not play such an important role since the intrinsic scatter of the EW distribution is much higher.
    \item We applied our model to a number of real-world data, providing examples of possible usage of the automated spectral classification in different environments. We included either cases of isolated massive stars or companion stars in HMXBs in our applications in order to assess the applicability of our classifier in range of science cases. We find that the model works well, especially for data of similar quality to the training sample. The code for the application of our models is available through the
    GitHub repository \footnote{\url{https://github.com/EliasKyritsis/SpectralClassifier}}.
\end{enumerate}

The main directions for improvement to increase the performance of our model are: a) the use of a larger sample with better statistics in the under-represented classes, as well as better representation of different metallicity populations; and (b) the use of the RF algorithm as a regression tool in order to determine stellar temperatures, which is the underlying parameter driving spectral types. 
A reliable tool for spectral classification, such as the one we developed in this work, is particularly important for the analysis of data from large surveys that target thousands of stars (or galaxies) and produce millions of spectra (e.g., OWN, NoMaDS, CAFE-BEANS, etc.). 
\section{Acknowledgements}
We thank the anonymous referee for his/her constructive comments and suggestions that helped to improve this work and the clarity of the manuscript. We thank Dr. Chris Evans and Dr. Philip Dufton for providing us the VFTS rectified data used in this work. In addition, we thank Dr. Itamar Reis for the useful comments regarding the PRF algorithm. EK acknowledges support from the Public Investments Program through a Matching Funds grant to the IA-FORTH. GM acknowledges funding from the European Research Council (ERC) under the European Union’s Horizon 2020 research and
innovation programme (grant agreement number 772086). The research leading to these results has received funding from the {\it European Research Council} under the European Union's {\it Seventh Framework Programme} (FP/2007-2013) / {\it ERC} Grant Agreement n.~617001, and the {\it European Union’s Horizon 2020} research and innovation programme under the {\it Marie Sk\l{}odowska-Curie RISE} action, Grant Agreement n.~691164 ({\it ASTROSTAT}). We wish to thank the "2019 Summer School for Astrostatistics in Crete" for providing training on the statistical methods adopted in this work. We also thank Jeff Andrews for organizing the SMAC (Statistical methods for Astrophysics in Crete) seminar series.

\bibliographystyle{aa}
\bibliography{references}

\begin{appendix} 

\section{An alternative bands-based method for measuring EW and its uncertainty}\label{bands-based}
Although the spectral fitting method (Sec. \ref{subsec:EWs_EWs_errors}) provides the most reliable measurements of the EW, it is computationally intensive and it often requires optimization for the particular data set it is applied to. For this reason, we also consider the more commonly used bands-based method. In this method the EW of a given line is obtained from measurement of the intensity in three bands: one centered on the spectral line of interest and two used to measure the continuum on the blue and the red sides of the spectral line, respectively. For the calculation of the EW, the basic formula in Equation \ref{EW_basic_formula} can be transformed as
\begin{equation}
\label{EW_formula}
\centering
EW = - \left( d\times N - \sum_{i=1}^{N}\frac{F_{line_{i}}}{F_{cont_{i}}}d \right) = - \left( d\times N- d\times\frac{1}{C}\sum_{i=1}^{N}F_{line_{i}} \right), \ 
\end{equation}
since in our case the spectral density is measured in individual pixels and their relation to the wavelength is given by the dispersion of the spectrum $(d)$.
The flux of the continuum is considered constant (C) over the wavelength range of N pixels of the line. Thus, we can measure the total flux of the spectral line as the sum of the pixel values included within the $\lambda_{1}$-$\lambda_{2}$ wavelength region. To estimate the continuum flux density at the location of the line, we interpolated between the flux density at the red and the blue continuum regions based on Equation \ref{Continuum_formula}:
\begin{equation}
\label{Continuum_formula}
\centering
C = C_{blue} + \frac{C_{red}-C_{blue}}{\lambda_{red}-\lambda_{blue}}(\lambda_{line}-\lambda_{blue}) \ ,
\end{equation}
where $C_{blue}$ and $C_{red}$ are the average values for the continuum flux density at the blue and red sides (in $\AA$/pix), and $\lambda_{blue}$, $\lambda_{red,}$ and $\lambda_{line}$ are the central wavelengths of the blue, red continuum, and the line regions, respectively. Appropriate wavelength ranges for the spectral lines and their continuum regions were defined after visual inspection to ensure that they did not include other lines or artifacts \citep{PhD_maravelias}. Next, we calculated the mean intensity at the center of each continuum side, and by following the above procedure, we calculated the continuum flux density at the central wavelength of each line. In addition, when we measured the average continuum flux densities, we applied a sigma clipping method with an accepting threshold of 1$\sigma$ and by iterating once. In this way, we measured a pure continuum region that was not biased from outliers (i.e., spectral lines or artefacts). As a final step, this mean continuum value was used as an input in Equation \ref{EW_formula} to measure the EW of each spectral line.

For the calculation of the EW uncertainty, we applied error propagation to Equation \ref{EW_formula}: 
\begin{equation}
\label{EW_error_formula}
\centering
\delta EW = \sqrt{\left(\frac{(\sum_{i=1}^{N}F_{line_{i}})\delta C}{C^{2}}\right)^2 + \left(\frac{ (\sum_{i=1}^{N}\delta F_{line_{i}}^{2})^{1/2}}{C}\right)^2} \ .
\end{equation}
The calculation of the total continuum uncertainty ($\delta C$) and the total flux uncertainty of the line ($\delta F_{line}$) depends on the availability of flux uncertainties in the initial spectral data. Thus, for the 2dF, VFTS, and GOSC data, from which the information of flux uncertainty was missing, the second term in Equation \ref{EW_error_formula} was equal to zero, and the continuum uncertainty was calculated from the standard error of the mean of the flux of the pixels in the continuum region:
\begin{equation}
\label{Continuum_error_formula}
\centering
\delta C = \sqrt{ \left(\frac{\sigma_{C_{blue}}}{\sqrt{N_{C_{blue}}}}\right)^2 + \left(\frac{\sigma_{C_{red}}}{\sqrt{N_{C_{red}}}}\right)^2} \ .
\end{equation}
On the other hand, for the LAMOST data that provide flux uncertainty for each spectral bin, the total flux uncertainty of the line was calculated as
\begin{equation}
 \delta F_{line}= \sqrt{\sum_{i=1}^{N}\delta F_{i}^{2}} \ ,  
\end{equation}
where $\delta F_{i}$ is the flux uncertainty of N-th pixel in the line. Finally, the total continuum uncertainty was calculated as
\begin{equation}
\centering
\delta C = \sqrt{ \left(\frac{\sigma_{\delta C_{blue}}}{\sqrt{N_{C_{blue}}}}\right)^2 + \left(\frac{\sigma_{\delta C_{red}}}{\sqrt{N_{C_{red}}}}\right)^2} \ ,
\end{equation}
where we used the standard deviation of the flux uncertainties of the blue and red continuum regions, respectively, instead of the standard deviation of the flux values that we used in Equation \ref{Continuum_error_formula}.
We applied this method to the entire data set of the spectra using the same spectral lines and corresponding bands presented in Sect. \ref{subsec:EWs_EWs_errors}.

\begin{figure*}
\centering
        \includegraphics[width=\textwidth]{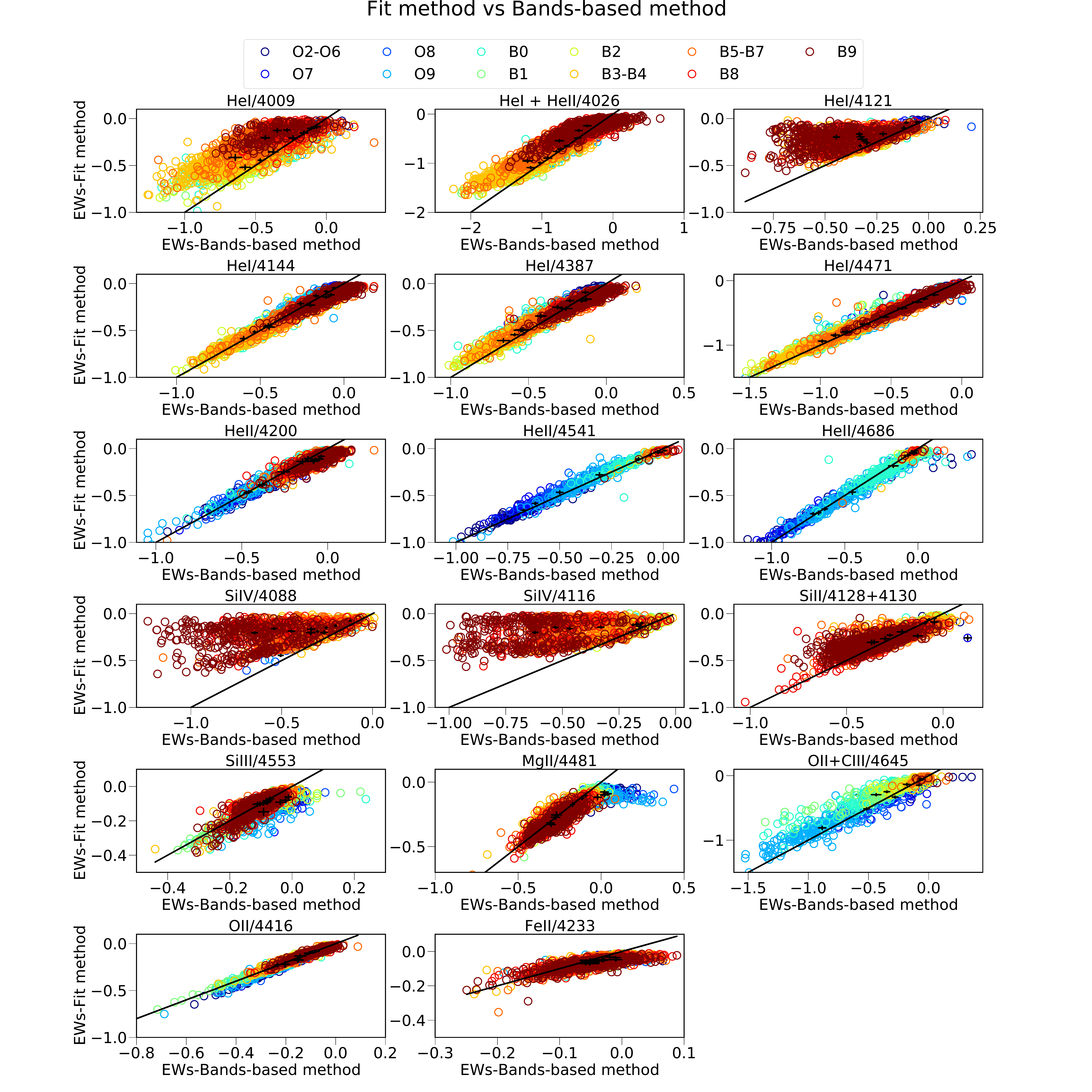}
        \caption{One-to-one comparison of the EW measurements per spectral line and per spectral class between the two methods. Black crosses depict the median EW uncertainty per spectral class.}
    \label{Fig.:ew_fit_band_comparison}
\end{figure*}

In Fig. \ref{Fig.:ew_fit_band_comparison}, we present the one-to-one comparison of the EW measurements per spectral line and per spectral class between the spectral fit and the bands method. With black crosses, we depict the median EW uncertainties per spectral class. As shown, the two methods are consistent mainly for the strongest spectral lines (e.g., \ion{He}{II}, \ion{He}{I}/4471 \AA , \ion{He}{I}/4144 \AA,  \, etc.). On the other hand, for weaker spectral lines (e.g., \ion{He}{I}/4009 \AA , \ion{He}{I} + \ion{He}{II}/4026 \AA, \, etc.) the bands-based method seems to overestimate the EW value, since in these cases the measurements are dominated by continuum noise. This is particularly prominent in the case of the \ion{Si}{IV}/4088 \AA \, and \ion{Si}{IV}/4116 \AA \,  spectral lines. These two lines are relatively weak and close to the H$\delta$ Balmer line, resulting in strong blending, especially for late B-type stars with strong Balmer lines. This effect makes the measurement of these two lines very unreliable because the bands-based method actually measures the broad H$\delta$ wings instead of the \ion{Si}{IV} spectral lines. However, this is solved with the spectral line fitting method presented in Sect. \ref{subsec:EWs_EWs_errors}. Regarding the EW uncertainties measured from both methods, we find that the fit method estimates the EW error slightly more accurately than the bands method. The most important thing, however, is that the scatter of the data is not driven by the uncertainties of the measurements (which are significantly lower), but from the intrinsic scatter of the EW values that hampers the separation of adjacent spectral classes based on the EW of characteristic spectral lines.

\begin{figure*}
        \includegraphics[width=\columnwidth]{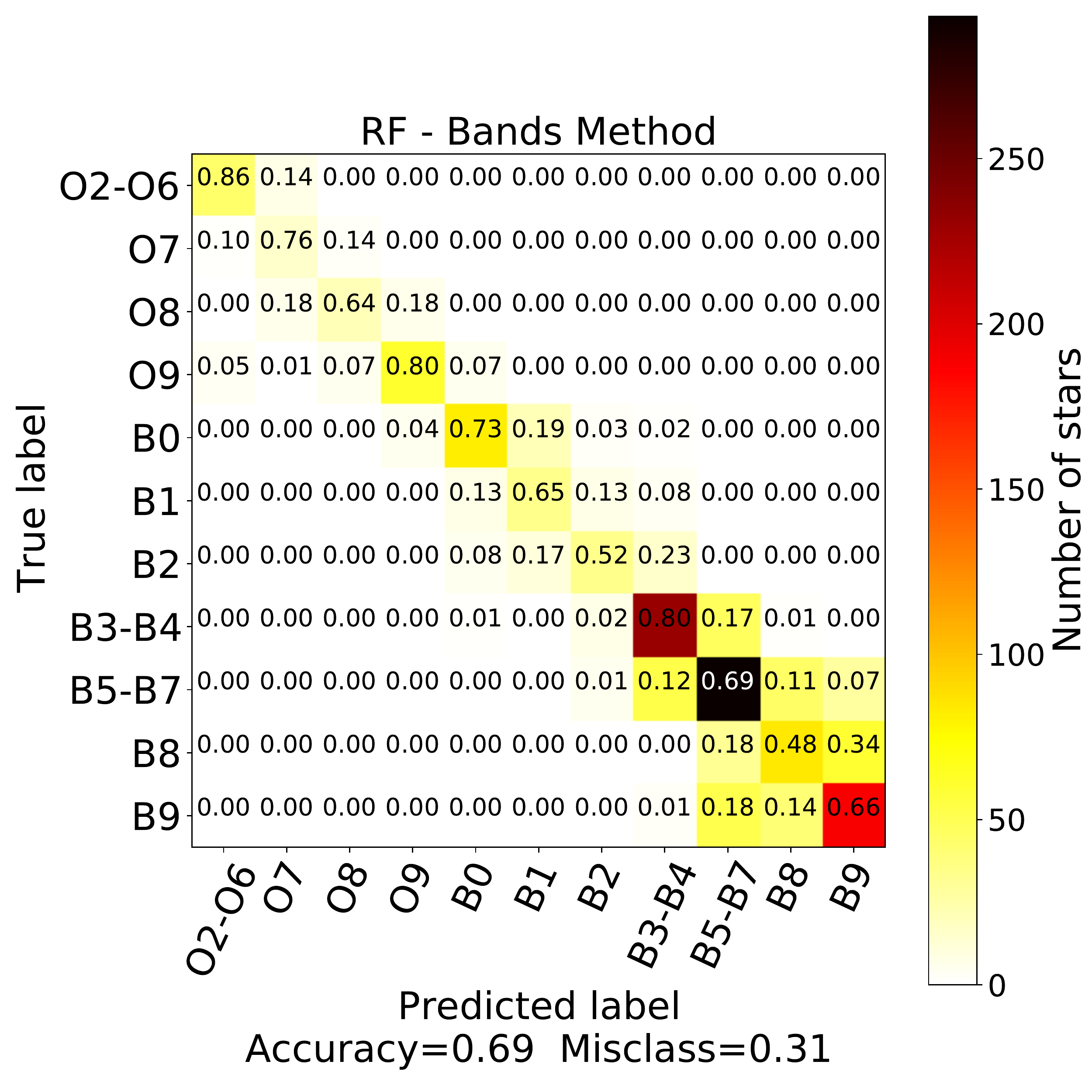}
        \includegraphics[width=\columnwidth]{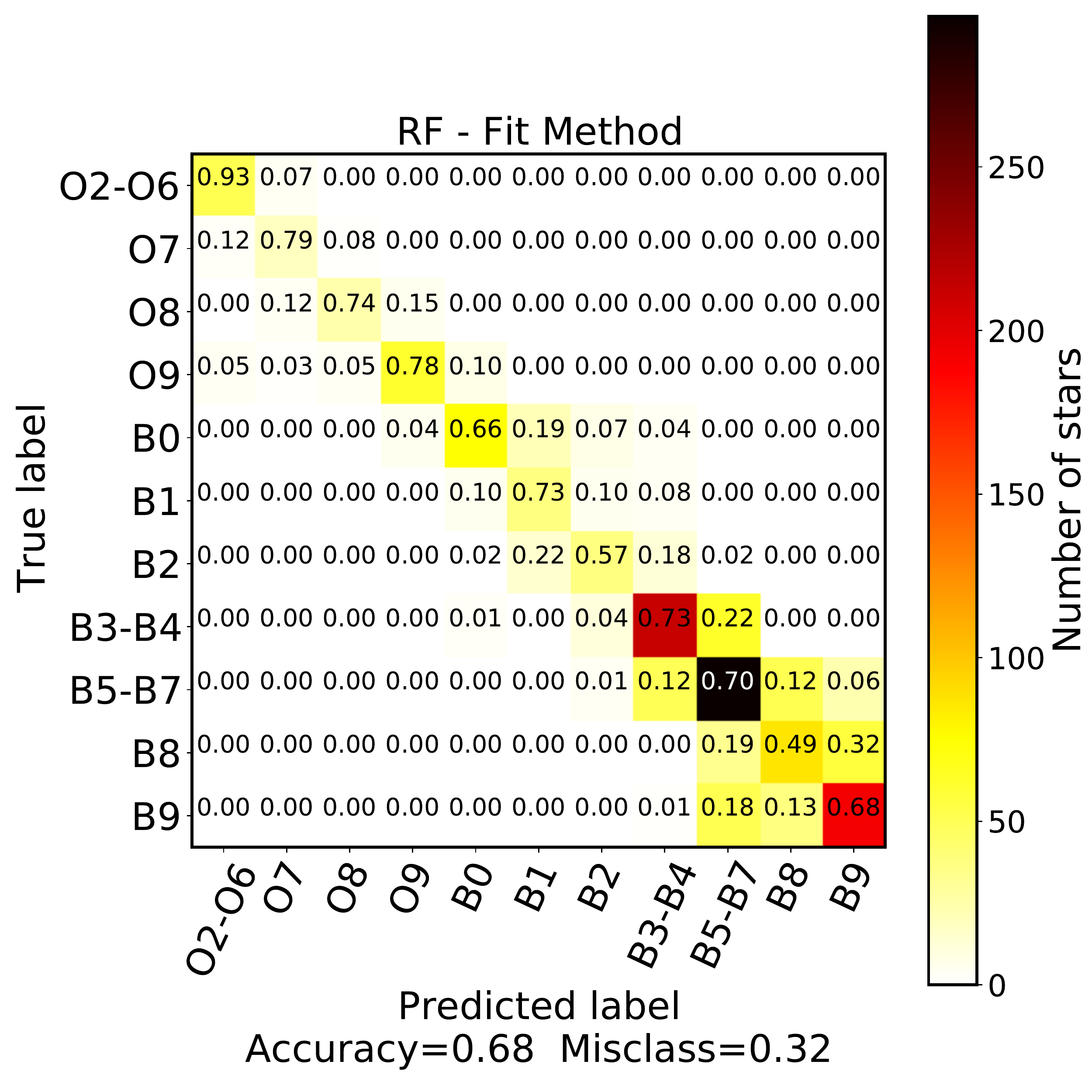}
    \caption{Top right panel shows the normalized confusion matrix of the best RF model trained on EW values obtained from the fit method, while the left panel shows the normalized confusion matrix of the RF best model trained on values obtained from bands-based method. The k-fold validation for both gives exactly the same CV score, 69$\pm$1$\%$, indicating that any failure of the RF algorithm to correctly predict all the spectral classes is not due to inaccurate measurements but due to the intrinsic scatter of the EW within this range of spectral classes.}
    \label{Fig.:Consusion matrix RF_fit_RF_bands}
\end{figure*}

To compare the performance of the RF algorithm, we trained it on the same data set but with EW measurements obtained from both methods. We used the 15 spectral lines that can be measured either from the fit method or from the bands-based method as features, and we excluded \ion{Si}{IV}/4088 \AA \, and \ion{Si}{IV}/4116 \AA , for which the bands-based method provides unreliable measurements. In Fig. \ref{Fig.:Consusion matrix RF_fit_RF_bands}, we present the results of this comparison. As shown from the confusion matrices of the two methods, they present similar behavior in terms of general accuracy, as well as in terms of the recall score per spectral class. In addition, the k-fold validation for both gives exactly the same CV score, 69$\pm$1$\%$. The consistency between the measurements of both methods is further proof that the score of the RF algorithm is driven by the intrinsic scatter of the EW values.

To conclude, we present an alternative method for measuring the EW of characteristic spectral lines. Although in weak spectral lines it tends to overestimate the EW, in general it demonstrates similar performance to the basic fit method we presented in the main paper. However, the fit method is of course the most accurate method that can be applied to such of problems. We suggest that potential users of our tool use this bands-based method if they are interested in the classification of large data sets, since the time needed to measure the EW with spectral fitting of the lines is significantly higher than with the bands-based method. 

\section{A reduced feature scheme for the application of the RF-bands method}\label{10-feature scheme}
Although in our analysis we considered the full scheme of lines, aiming to receive the maximum performance of our model per spectral class, in Table \ref{tab:Append_10_spec_lines} we present an alternative scheme that can also used for the spectral classification of OB stars in green. In particular, it comprises the ten most important spectral features resulting from the SFFS analysis in Sect. \ref{subsec:features-opt}. This reduced scheme includes the strong spectral lines \ion{He}{I}/4471 and \ion{Mg}{II}/4481, the most important \ion{He}{II} lines (indicators of O-type stars), as well as \ion{He}{I} lines, which are good indicators of the early- and mid-B-type stars, while the weak \ion{Si}{IV}/4088-4116 are not included. The absence of \ion{Si}{IV} lines is particularly important because one can use the bands-based method, which, as we discussed in Appendix  \ref{bands-based}, overestimates their EW measurements. Furthermore, given the comparison between the bands-based and the fit method for the reduced scheme (see Fig.\ref{Fig.:ew_fit_band_comparison}), we would expect the bands-based EW measurements to yield a score similar to $\sim 68 \%$, as is suggested by the SFFS algorithm for the fit method.
The overall performance of $\sim 68 \%$ suggests that it can be applied to large data sets with lower S/Ns ($\sim 20-50$), where the measurement of weaker lines is not reliable.    

\begin{table}
\centering
\caption{Reduced classification scheme, which can be measured with the bands-based method.}
\begin{tabular}{cc}
\hline\hline
Line ID & $\lambda_{\textrm{central}}$ \\
        & ({\AA})                            \\\hline
\rowcolor{green}\ion{He}{I}     & 4009 \\
\rowcolor{green}\ion{He}{I}+\ion{He}{II}& 4026 \\
\ion{Si}{IV} &4088 \\
\ion{Si}{IV} &4116 \\
\ion{He}{I}  & 4121 \\
\rowcolor{green}\ion{Si}{II}    & 4128+4130 \\
\ion{He}{I} & 4144 \\
\rowcolor{green}\ion{He}{II}    & 4200 \\
\ion{Fe}{II}    & 4233 \\
\rowcolor{green}\ion{He}{I}     & 4387 \\
\rowcolor{green}\ion{O}{II}     & 4416 \\
\rowcolor{green}\ion{He}{I}    & 4471 \\
\rowcolor{green}\ion{Mg}{II}    & 4481 \\
\ion{He}{II}    & 4541 \\
\rowcolor{green}\ion{Si}{III}   & 4553 \\
\ion{O}{II}+\ion{C}{III}& 4645 \\
\rowcolor{green}\ion{He}{II}    & 4686 \\
\hline
\end{tabular}
\label{tab:Append_10_spec_lines}
\end{table}

\end{appendix}

\end{document}